\documentclass[preprint,amsmath,amssymb,superscriptaddress,notitlepage]{revtex4-2}
\usepackage{graphicx}
\usepackage[utf8]{inputenc}
\usepackage{ulem}
\usepackage{amssymb}
\usepackage{color}
\usepackage{units}
\usepackage{float}
\usepackage{amsmath}
\usepackage{url}
\usepackage{hyperref}
\usepackage{epsfig}
\usepackage{bm}
\usepackage{xspace} 
\hypersetup{
  colorlinks = true, 
  urlcolor   = blue,
  linkcolor  = red, 
  citecolor  = blue 
}

\begin{document}
\title{Magneto-elasto-resistivity in FeSe}
\author{M.~Wissmann$^{1,2,3}$, L.~Fanfarillo$^{4}$, X.-C.~Hong$^{1}$, S.~Caprara$^{5}$, S.~Aswartham$^{1}$,  B.~Büchner$^{1,2}$, C.~Hess$^{1,2,7}$,  G.~Seibold$^{6}$ and F.~Caglieris$^{8}$\\ 
$^{1}$\textit{IFW Dresden, P.O. Box 270116, 01171 Dresden, Germany} \\
$^{2}$\textit{Institut f{\"u}r Festk{\"o}rper- und Materialphysik, Technische Universit{\"a}t Dresden, 01062 Dresden, Germany }\\
$^{3}$\textit{Université Grenoble Alpes, CNRS, CEA, Grenoble-INP, Spintec, 38000 Grenoble, France}
$^{4}$\textit {Istituto dei Sistemi Complessi (ISC-CNR), Via dei Taurini 19, I-00185 Rome, Italy}\\
$^{5}$\textit{Dipartimento di Fisica, Sapienza Universit\`a  di Roma, Piazzale Aldo Moro 5, I-00185 Rome, Italy}\\
$^{6}$\textit{Institut f\"ur Physik, Brandenburg Technical University Cottbus-Senftenberg, 03013 Cottbus, Germany}\\
$^{7}$\textit{Fakultät für Mathematik und
Naturwissenschaften, Bergische Universität Wuppertal, Gaußstraße 20, 42119 Wuppertal, Germany}\\
$^{8}$\textit{CNR-SPIN, Corso Perrone 24, 16152 Genova, Italy}
}

\begin{abstract}
FeSe stands out among iron-based superconductors due to its extended nematic phase without the onset of long-range magnetic order. While strain-dependent electrical resistivity has been extensively explored to probe nematicity, its influence on magneto-transport properties remains less understood. In this work, we present measurements of the magneto-elasto-resistivity in FeSe as a function of temperature and applied magnetic field. Using a minimal multiband Boltzmann model for transport we derive analytical expressions that capture the magnetic behavior of the whole set of experimental data both in the paramagnetic and in the nematic phase. These findings indicate that a multiband framework can robustly describe the magneto-elasto-transport properties in FeSe and arguably in other iron-based superconductors.
\end{abstract}

\maketitle

\section{Introduction}


Since their discovery, iron-based superconductors (Fe-SCs) have provided a rich platform to explore the interplay between structural, magnetic, and orbital degrees of freedom in unconventional superconductivity\cite{Nature2022}. 
Among the various emergent phenomena, the so-called nematic phase, defined by the spontaneous breaking of the crystal rotational symmetry while preserving translational invariance, has attracted particular attention due to its close connection with magnetism and superconductivity \cite{NatPhys2014, Qisi_Frontiers2022, Bohmer_NatPhys2022}.
In Fe-SCs, nematicity typically emerges before or alongside magnetic ordering, suggesting a close interplay between the two. Nematic fluctuations can enhance magnetic interactions, while magnetic fluctuations may contribute to the onset of nematicity and their mutual influence is believed to play a role in shaping unconventional superconductivity \cite{NatPhys2014,Qisi_Frontiers2022, Bohmer_NatPhys2022}.
In this context, the FeSe compound stands out as a notable exception. It exhibits an extended nematic phase below $T_S \approx 90$ K without developing long-range magnetic order, and hosts bulk superconductivity below a critical temperature $T_C \approx 9$ K \cite{Bohmer_JPCM2018, Kreisel_Symmetry2020}. Although the absence of long-range magnetism does not preclude the presence of magnetic fluctuations, FeSe offers a unique opportunity to investigate the relationship between nematicity and superconductivity in the absence of magnetic order.
There is nowadays a general agreement on the electronic nature of the nematic state, however the exact origin of nematicity in FeSe is still under debate. The lack of magnetic order motivated a scenario based on orbital or charge ordering \cite{Baek_NatMat2014, Su_JPCM2015, Massat_PNAS2016}, while the detection of sizeable spin fluctuations \cite{Rahn_PRB2015, Wang_NatMat2016, Chen_NatMat2019, Zhou_npjQM2020} triggered an intense investigation on the interplay between spin and orbital degrees of freedom \cite{Fanfarillo_PRB2015, Glasbrenner_NatPhys2015, Christensen_PRB2016, Fernandes_Rep2017, Chen_NatMat2019, Kreisel_Frontier2022}, leading to alternative picture of nematicity driven by orbital-selective spin fluctuation \cite{Fanfarillo_PRB2016, Fanfarillo_PRB2018, Kreisel_Symmetry2020}.\\
A powerful experimental technique to probe the coupling between nematic and electronic degrees of freedom is the measurement of the strain dependent electrical resistivity, or elasto-resistivity (ER), which has been widely used over the past decade \cite{ChuScience, KuoScience, PhysRevB.106.054508, 122XC, FeSePhysRevLett.117.127001, FeSePhysRevX.11.021038, PhysRevLett.125.067001, NpjCcaglieris}. In particular, the measurement of the ER as a function of temperature and doping allowed to identify the electronic origin of nematicity in Fe-SCs and to reveal a correlation between the enhanced nematic fluctuations and the increase of the superconducting critical temperature at the optimal doping. Curiously, much less attention has been devoted to the study of the magneto-elasto-resistivity (MER), the measurement of ER under an applied magnetic field, which can offer complementary insights into the symmetry-breaking mechanisms by probing the magnetic field dependence of strain-induced transport anisotropy.
To address this gap, we present a detailed investigation of MER in FeSe as a function of temperature and applied magnetic field. We analyze the experimental data using the Boltzmann transport theory, taking into account the multiband nature of the electronic structure of FeSe and the transport anisotropy induced by nematicity below $T_S$. 
While the effects of electronic correlations and bands' orbital composition are not explicitly modeled, they are effectively incorporated at the microscopic level in the $x/y$ band conductivities, used in defining the fitting parameters.
Our analysis captures the essential behavior observed in the magneto-transport experiments. Notably, two key features of the MER data emerge directly from the analytical expressions, without any fine-tuning of fitting parameters: (i) the field-independent behavior above $T_S$ for out-of-plane (i.e., parallel to the crystallographic $c$-axis) fields, and (ii) the absence of magnetic-field dependence for in-plane (i.e., perpendicular to the crystallographic $c$-axis) fields.
These results demonstrate that a minimal two-band model robustly captures the essential magneto-elasto-transport features in FeSe, suggesting a unified framework for analyzing MER in other Fe-SCs.

\section{Experimental methods}

FeSe single crystals have been synthesized according to the procedure reported in Ref.\,\cite{Baek_2020}. The MER measurement setup has been described in Ref.\,\cite{PhysRevLett.125.067001, 122XC}. All the measurements have been performed using home-made probes inserted in an Oxford Instruments cryostat, equipped with a 15/17\,T magnet. The transport characterization has been carried out adopting a standard 4-wire configuration, in which the electrodes consist of 50\,$\mu$m-thick silver wires attached to the sample with silver paints. To apply the uniaxial strain, the sample has been glued on a commercial piezoelectric actuator endowed with a low-temperature coating to perform measurements in cryogenic environment. The uniaxial strain has been applied along the 110 crystalline direction, i.e. activating the B$_{2g}$ mode, associated to the nematic fluctuations. The magnetic field has been oriented both out of plane and in plane (i.e., parallel and perpendicular to the crystallographic $c$-axis).

\begin{figure}[tbh]
\includegraphics[width=0.85\columnwidth]{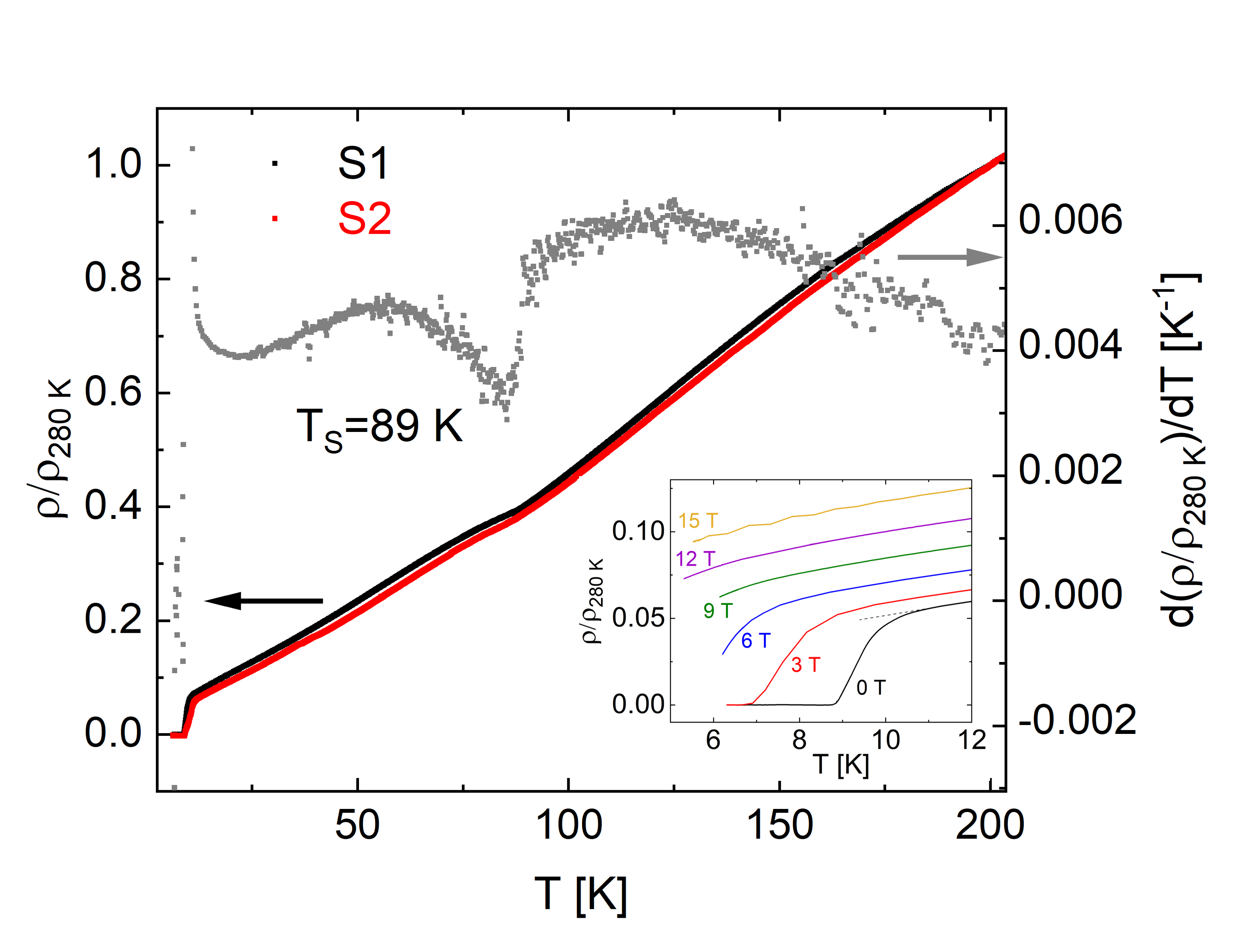}
\caption{Temperature dependence of the resistivity of samples S1 and S2 normalized to the values at 280\,K. The temperature-derivative is shown in gray and clearly identifies $T_S$ and $T_C$. Inset: Temperature dependence of the resistivity of sample S1 across the superconducting transition at several magnetic fields. \label{Figure1}}
\end{figure}

\section{Results}
\begin{figure}[!t]
\includegraphics[width=1.03\columnwidth]{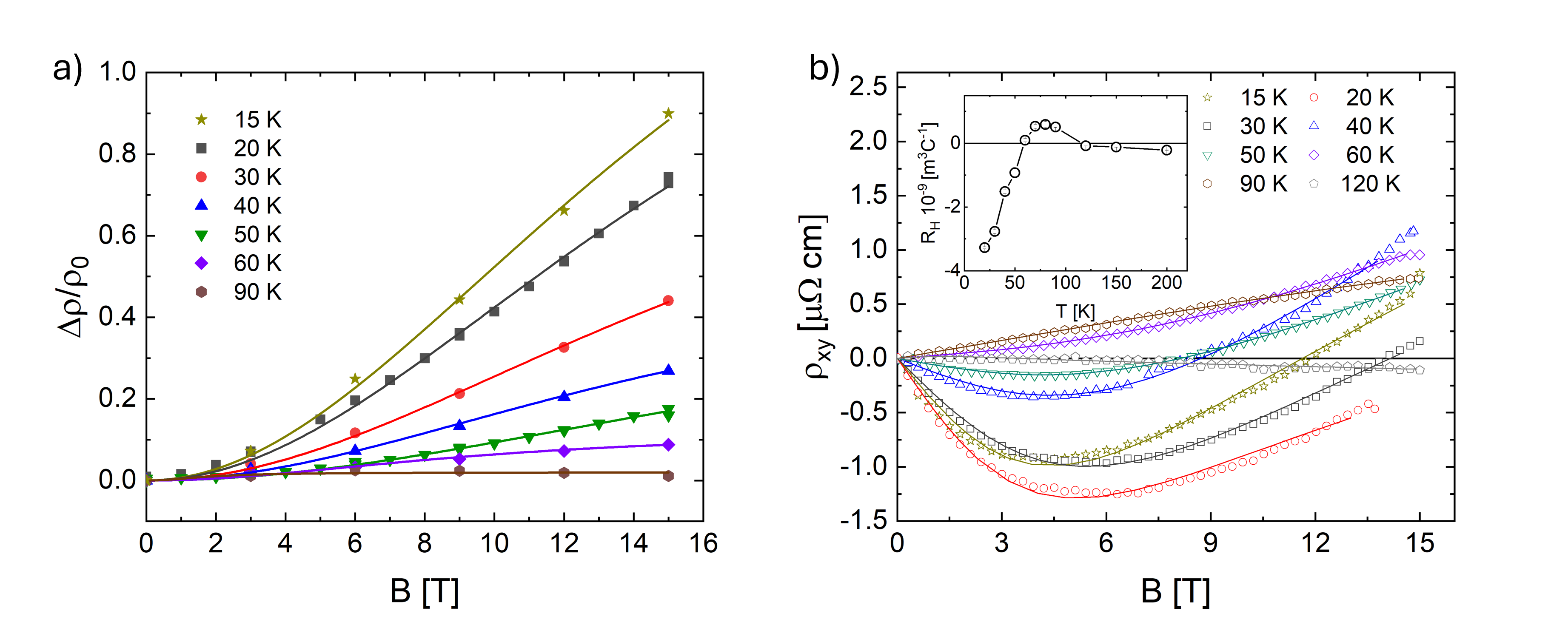}
\caption{(a) MR of sample S1 as a function of the applied magnetic field $B$, for different temperatures. (b) Hall resistivity $\rho^{xy}$ vs $B$ of samples S1, measured at different temperatures. Inset: Hall-coefficient $R^H_0$, here derived as the slope of the linear fits in the low-$B$ limit . In both panels (a)-(b), the theoretical fit using Eq.s\,(\ref{eq:magneto}) and (\ref{eq:RH}) are shown as solid lines on top of the experimental data (symbols). Fitting parameters are reported in App.\,\ref{App:Fit}.
\label{Figure2}}
\end{figure}

The temperature-dependent resistivity (normalized to the value at $T$=280 K) $\rho/\rho_{280\ K}$ of two mono-crystalline FeSe samples (S1 and S2) from the same batch is displayed in Fig.\,\ref{Figure1}. The two curves almost overlap and exhibit the characteristic signatures of the structural and superconducting transitions at $T_S= 89$\,K and $T_C=9$\,K, respectively, in agreement with previous reports \cite{doi:10.1073/pnas.1413477111,Bohmer_JPCM2018}. Fig.\,\ref{Figure2} shows the results of magneto-transport measurements of sample S1. In particular, Fig.\,\ref{Figure2}(a) presents the magnetoresistivity (MR) $\frac{\Delta \rho(B)}{\rho(0)}\equiv\frac{\rho(B)-\rho(0)}{\rho(0)}$ as a function of the magnetic field $B$, applied along the crystallographic $c$-axis, at several temperatures. The MR is relatively small in the tetragonal phase and it grows below $T_S$, reaching the remarkable value of 90$\%$ at 15\,K and 15\,T, without signatures of saturation. 

The field-dependence of the Hall resistivity $\rho^{xy}$ is shown in Fig.\,\ref{Figure2}(b). 
Above $T_S$, $\rho^{xy}$ is linear-in-field and negative, while at $T=$ 90 K $\rho^{xy}$ turns positive with a larger slope. With further decreasing the temperature, in the nematic phase, $\rho^{xy}$ becomes strongly non linear, with field-induced sign changes. The inset shows the temperature dependence of the Hall coefficient $R^H_0$, thereby evaluated as $R^H_0=\frac{d\rho^{xy}}{dB}|_{B\rightarrow 0}$, which exhibits a first sign change at $T \approx 100$\,K and a second sign change at $T \approx 60$\,K, in agreement with previous reports \cite{PhysRevResearch.3.023069, Caglieris01102012}. 

\begin{figure}[!t]
\includegraphics[width=1.03\columnwidth]{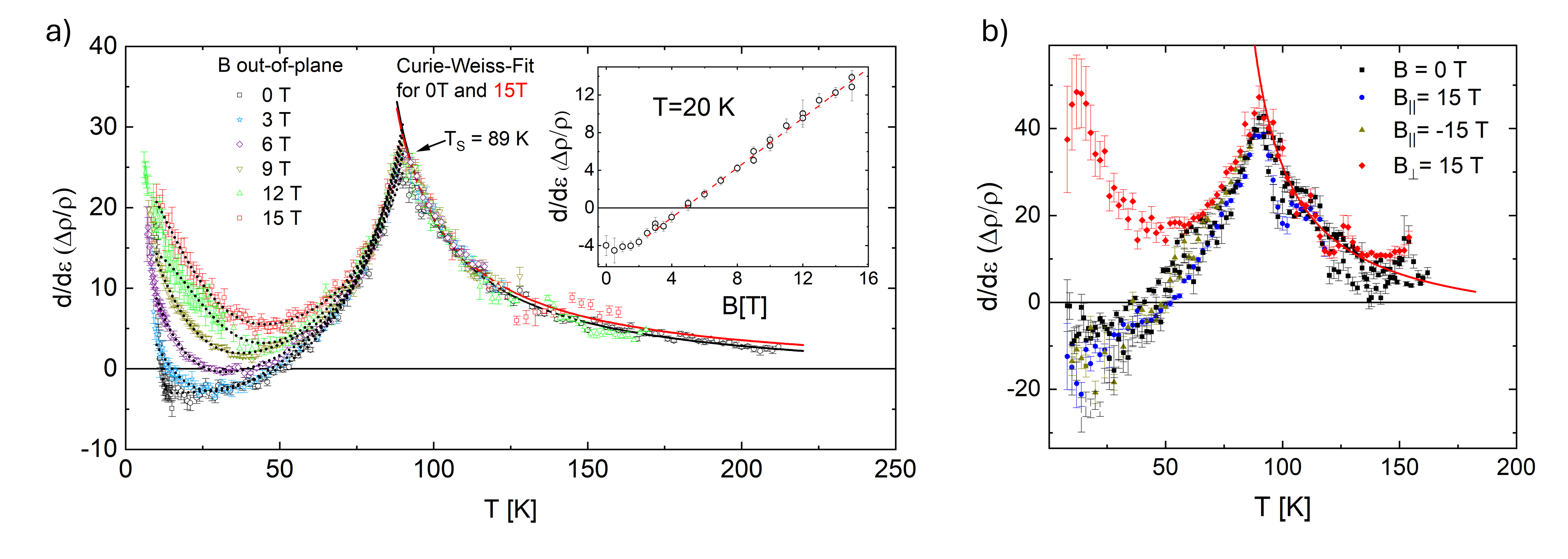}
\caption{(a) ER on sample S1 in out-of-plane fields, from 0\,T (black) to 15\,T (red), in 3\,T-steps shows a systematic increase with increasing field in the orthorhombic phase. The Curie-Weiss-like behavior in the tetragonal phase, including the divergence temperature $T_S$, is field-independent. The theoretical fits using Eq.\,(\ref{eq:elasto}) are shown as dashed lines on top of the experimental data (symbols). Fitting parameters are reported in App.\,\ref{App:Fit}. (b) MER of sample S2 for different magnetic field configurations: zero-field (black), 15\,T and $-15$\,T in-plane (blue and brown) and 15\,T out-of-plane (red). No effect of magnetic fields is visible for in-plane fields.\label{Figure3}}
\end{figure}

Fig.\,\ref{Figure3}(a) presents the MER of S1 as a function of the temperature for different magnetic fields applied along the crystallographic $c$-axis. The zero-field ER is consistent with existing literature. Above $T_S$, ER is positive and exhibits the characteristic Curie-Weiss-like behavior, generally interpreted as the contribution of electron-driven nematic fluctuations. The extracted Curie-Weiss temperature is $T^*\approx 71$\,K. Below $T_S$, the ER drops and becomes negative for $T<50$\,K.

The application of a magnetic field has almost no effect above $T_S$ and the Curie-Weiss fit of the MER at 15\,T returns the value of $T^*=70$\,K, in good agreement with the one obtained from the fitting of the zero-field ER. Below $T_S$, the MER strongly increases towards positive values and shows a pronounced upturn with decreasing $T$. The magnetic field dependence of the MER at $T=20$\,K and 50\,K is substantially linear for $B>4$\,T, with a departure from linearity at lower fields. 

Fig.\,\ref{Figure3}(b) shows the MER of S2 with $B$=0 and 15\,T, applied both parallel and perpendicular to the crystallographic $c$-axis of the sample. Both the ER and the MER at 15\,T, parallel to the $c$-axis, are in good agreement with the measurement on sample S1. In addition, in this second set of data we verified that a magnetic field oriented in plane has no effect on the MER even below $T_S$, with the MER curves at $\pm$15\,T overlapping over the zero-field ER.

\section{Discussion}

Our goal is to derive a minimal theoretical framework that consistently describes the full set of magneto-elasto-resistivity (MER) measurements presented for FeSe. A natural starting point for modeling electron transport is the Boltzmann equation, which provides a semi-classical framework for describing charge conduction. In Fe-based superconductors, this approach must be extended to a multiband formulation to account for the complex electronic structure that governs their transport properties.
To keep the discussion as simple and transparent as possible, we adopt a minimal multiband model consisting of two bands. The onset of nematic order below the structural transition temperature $T_S$ is incorporated by introducing an $x/y$ anisotropy in the band conductivities. Importantly, we do not make any assumptions about the microscopic origin of nematicity. To preserve generality, we work in the band basis without explicitly introducing orbital degrees of freedom.
As we will show, this minimal model is sufficient to reproduce the key features of the MER data without invoking orbital-selective effects. While the orbital composition of the bands is undoubtedly important at the microscopic level, particularly in determining the anisotropic conductivities in the nematic phase \cite{Fernandez_PRB2019, Marciani_PRB2022}, these effects are effectively absorbed into the fitting parameters that define the $x/y$ band conductivities in our model.

In what follows, we summarize the main steps of the derivation and report only the equations relevant for the discussion. A detailed derivation of the Boltzmann transport theory for a two-band system in both the paramagnetic and nematic phases is provided in App. \ref{App:Boltzmann}.

Assuming that the electronic scattering rate is independent of the magnetic field, the resistivity tensor in the presence of an out-of-plane magnetic field (denoted as $B_z$) takes the form:
\begin{equation}
\label{eq:rho}
  \hat{\rho} =
  \left(
  \begin{array}{cc}
    \rho^x & B_z R^H \\
    -B_z R^H & \rho^y
  \end{array}\right)
  \,,
\end{equation}
where $\rho^{x/y}$ and $\rho^{xy} = R^H B_z$ are expressed in terms of the components of the band resistivity tensors $\hat{\rho}_n$ ($n = 1, 2$ is the band index). The explicit expressions for $\rho^{x/y}$ and $R^H$ in the two-band model are given in Eq.s (\ref{eq:rx2})–(\ref{eq:rb2}).
The Hall coefficient is given by:
\begin{equation}\label{eq:RH}
  R^H=\frac{a_0 + a_1 B_z^2}{1+a_2 B_z^2}\,,
\end{equation}
where $a_0$, $a_1$, and $a_2$ are functions of the components of $\hat{\rho}_n$. This expression holds in both isotropic and anisotropic cases, only the explicit forms of $a_0$, $a_1$, and $a_2$ differ, as detailed in App. \ref{App:Boltzmann}.
Similarly, the magnetoresistance (MR) can be written as:
\begin{equation}\label{eq:magneto}
\frac{\Delta \rho(B)}{\rho(0)} =\frac{1+b_0 B_z^2}{1+b_1 B_z^2}-1\,,
\end{equation}
where $b_0$ and $b_1$ are also functions of the band resistivity tensors, with their explicit forms in the isotropic and anisotropic case provided in App. \ref{App:Boltzmann}.

To analyze the MR data, we treat $a_{0,1,2}$ and $b_{0,1}$ as fitting parameters. The results of the fitting procedure are shown in Fig. \ref{Figure2}(a)-(b), alongside the experimental data. The corresponding parameter values are listed in App. \ref{App:Fit}. A good fit is obtained using a physically reasonable set of parameters.
It is worth noting that the fit assigns different values to $b_1$ and $a_2$, although within our model these parameters are defined by the same expression (see App. \ref{App:Boltzmann}). We attempted a simultaneous fit of MR and $\rho^{xy}$ under the constraint $b_1 = a_2$. As expected, the inclusion of an additional constraint slightly worsens the fit quality, particularly at low temperatures. However, the overall trend of the fitting parameters remains unchanged (see App. \ref{App:Fit}), which supports the internal consistency of our minimal two-band model in capturing the essential features of the magneto-elasto-transport response in FeSe.
Importantly, since the parameters $b_1$ and $a_2$ primarily affect the MR and $\rho^{xy}$ in the high-field regime, their discrepancy may point to a limitation of the standard perturbative treatment of the Boltzmann equation via the Jones-Zener expansion \cite{jonesZener34}. This approximation assumes the scattering rate to be independent of the magnetic field. A more rigorous solution  of the Boltzmann equation, as discussed in \cite{Chambers_ProcPhysSocA1952}, accounts for the magnetic field’s influence on the time evolution of the wave vector, thereby modifying the scattering rate along the trajectory. In such a framework, $b_1$ and $a_2$ are no longer expected to be equal, however, to undertake this approach one would be forced to abandon the analytical treatment in favor of a fully numerical analysis.

The resistivity anisotropy is defined as $N = 2 (\rho^x-\rho^y)/(\rho^x+\rho^y)$, where the specific forms of $\rho^{x/y}$ for the two-band model are given in Eq.s (\ref{eq:rx2})-(\ref{eq:ry2}). To model the strain, we assume $ \rho_n^{x/y}=\rho_n^{0} \pm \frac{1}{2}\gamma_n \epsilon $, where $n=1,2$ is the band index and $\epsilon$ is the dimensionless strain. This corresponds to the simplest assumption that the resistivity anisotropy for each band is proportional to the strain only (i.e., $\rho_n^0$ is isotropic) and symmetric with respect to $x/y$. In this case, the MER reads 
\begin{equation}
\label{eq:elasto}
\frac{\partial N}{\partial \epsilon}=c_0\frac{1+c_1 B_z^2}{1 + c_2 B_z^2}\,,
\end{equation}
where $c_0$, $c_1$, $c_2$ are functions of the components of the band resistivity tensors $\hat{\rho}_n$, and of the strain coupling coefficients $\gamma_n$, whose explicit expressions are given in App.\,\ref{App:Boltzmann}. The analytical expression in Eq. (\ref{eq:elasto}) successfully captures the experimental behavior of MER in out-of-plane field above $T_S$. In particular, in the isotropic case and assuming $\gamma_1 \sim \gamma_2$, we find $c_1\sim c_2$ and $c_0 \sim \gamma/\rho$, implying that the MER becomes independent of the magnetic field above $T_S$, consistent with the experimental observation (see Fig. \ref{Figure3}(a) and discussion below Eq. (\ref{eq:ER_general})).

We further analyze the MER data using $c_{0,1,2}$ as fitting parameters. The result of the procedure is shown in Fig.\,\ref{Figure3}(a), together with the experimental data, while the fitting parameters are provided in App.\,\ref{App:Fit}. A good agreement with the experimental data is obtained using a set of parameters consistent with those used to analyze the MR data.

So far, we considered the transport properties of the system in the presence of an out-of-plane magnetic field $B_z$, as described by Eq.s (\ref{eq:rho})–(\ref{eq:elasto}). In Fig\,\ref{Figure3}(b) we show the experimental results obtained under an in-plane magnetic field and find that, in this case, the MER is independent of the magnetic field at all temperatures. To theoretically analyze this situation we derive the Boltzmann equation in the presence of an in-plane magnetic field.  The derivation follows the same steps as for the $B_z$ case. However, if we assume that the system is nearly 2-dimensional (with a larger effective mass along the $c$ crystallografic axis), the resistivity tensor in Eq.(\ref{eq:rho}) becomes nearly diagonal due to the negligible orbital-magnetic response (see App. \ref{App:Boltzmann}). As a consequence, the magnetic dependence of the MER, as expressed in Eq.\,(\ref{eq:elasto}),  is governed by almost vanishing $c_1$ and $c_2$ coefficients. This leads to a MER that is independent of the in-plane magnetic field at all temperatures, in agreement with the experimental findings.\\

Summarizing the theoretical analysis, we have shown that the magneto-elasto-transport measurements in FeSe can be simultaneously and consistently described using the Boltzmann equation within a minimal two-band model. The nematic order is incorporated phenomenologically by introducing an explicit $x/y$ anisotropy in the band resistivities below $T_S$. Remarkably, while the orbital composition of the electronic bands is certainly important for the microscopic determination of $\rho^{x/y}_n$ in the nematic phase, it is not a crucial ingredient for achieving a consistent description within the present framework. Working with a minimal band model, without introducing additional degrees of freedom, allows for an analytical treatment of the relevant physical quantities and provides clear physical insight.  In particular, key features of the MER magnetic-field dependence emerge naturally from inspection of the analytical expressions, without requiring fine-tuning of the fitting parameters. Finally, using a limited number of parameters, we could succesfully fit the full set of experimental data.

We emphasize that the present model focuses on describing the magnetic-field dependence of the magneto-elasto-transport measurements, while the temperature dependence of the various quantities is treated phenomenologically. Specifically, we account for the $x/y$ anisotropy of the band resistivities $\rho^{x/y}_n$ in the nematic phase, but to retain generality and preserve analytical tractability, we do not make any assumptions about the microscopic origin of nematicity. To theoretically describe the temperature dependence of MER near the nematic transition, including the expected $\gamma \sim 1/(T - T_S)$ behavior, one would need to define how electrons couple the nematic order parameter to the magnetic field within the model. However, the experimental behavior of MER around $T_S$, including the observation that $T_S$ remains unchanged under magnetic field, suggests that this coupling is weak and can be neglected to first approximation.
 
The superconducting critical temperature $T_C$ is the other relevant energy scale that should be incorporated to fully capture the temperature dependence of MER. In particular, the pronounced upturn of MER near $T_C$ in Fig. \ref{Figure3}(a) is consistent with a significant contribution from superconducting fluctuations to transport \cite{Fanfarillo_PRB2009, Fanfarillo_SST2014}. This contribution is also evident in the temperature dependence of the resistivity (inset of Fig. \ref{Figure1}), where a downturn in $\rho/\rho_{280,K}$ begins around $T \approx 11$,K, well above $T_C$. As expected, this fluctuation-induced contribution is suppressed with increasing magnetic field \cite{Dorin_PRB1993}. A detailed analysis of the regime near $T_C$ will be the subject of future work.

As a final remark, we notice that, although in this work we did not address the temperature dependence of the experimental quantities, the fitting of MR, Hall resistivity and MER at different temperatures provides valuable information on the temperature evolution of the fitting parameters and therefore can be used in future studies e..g. to identify the relevant scattering mechanisms within a more microscopic framework.

\section{Conclusion}

In this work we examined simultaneously the transport properties in magnetic field of the iron based superconductor FeSe, across the nematic transition. Beside the conventional magnetotransport (Hall effect and magnetoresistance), we explored the magneto-elasto-resistence, namely the elasto-resistance measured in magnetic field, which has been rarely experimentally addressed.

The overall behavior of the Hall and magnetoresistance data with respect to the magnetic field appears to be rather conventional and in agreement with previous reports \cite{Ovchenkov_JMMM2018}. However, a clear change appears at the structural transition temperature, below which both Hall and magnetoresistance display a much stronger field dependence. This phenomenology is compatible with a  modeling within the standard Boltzmann theory of transport, in which the multiband character of the electron structure, as well as the $x/y$ anisotropy in the nematic phase, are taken into account. Within this theoretical frame, analytical expression for both these experimental quantities can be derived. A good fit of the data is obtained in both cases using a reasonable and qualitatively consistent set of parameters. A better quantitative agreement would require the inclusion of the effect of the magnetic field on the scattering rate with the consequent need for a numerical solution of the Boltzmann equation. 

The magneto-elasto-resistence measured in magnetic field parallel to the crystallographic $c$-axis is independent of the magnetic field at any temperature. On the contrary, the magneto-elasto-resistence with $B_z$ displays a clear change at the structural transition temperature, with the absence of a magnetic field dependence above $T_S$, and a more complex behavior in the nematic phase. Remarkably the two-band Boltzmann model analytically captures the independence of the magneto-elasto-resistence from the magnetic field, both for in-plane fields at any temperature, and for out-of-plane fields above $T_S$. A good fit of the data can be obtained using a set of parameter qualitatively consistent with the ones used in the fit analysis of the magneto-resistence and Hall resistivity.

We would like to emphasize that orbital effects are usually needed to obtain a consistent picture of the transport in the nematic phase of iron-based materials, however in our magneto-elasto-transport measurements a multiband theoretical modeling appears adequate to successfully reproduce important features of the magnetic behavior of three different measured quantities. A confirmation to our hypothesis would come from the experimental analysis of other iron-based materials, exhibiting a different orbital nesting properties (e.g., BaFe$_2$As$_2$), where we expect to find similar features in the magneto-elasto-transport.

\appendix
\section{Multiband Boltzmann Theory}
\label{App:Boltzmann}

Here, we derive the Boltzmann equations for a multiband system in which, due to nematicity, the conductivities along the $x-$ and $y-$direction can be different. Our starting point is the linear response equation for each component $n$
\begin{equation}\label{eq:sigma}
\vec{J}_n=\hat{\sigma}_n \vec{E} \quad \quad  \quad \text{with} \quad \quad \quad
  \hat{\sigma}_n=\left(
  \begin{array}{cc}
    \sigma^{xx}_n & \sigma^{xy}_n \\
    \sigma^{yx}_n & \sigma^{yy}_n
  \end{array} \right),
\end{equation}
where $\vec{J}_n$ is the current density of the $n$-th band, $\vec{E}$ is the electric field, and $\hat{\sigma}_n$ is the conductivity tensor of the $n$-th band with components
\begin{eqnarray*}  
\sigma^{\alpha\beta}_n \  &=& \sigma^{\alpha\beta,(0)}_n+\sigma^{\alpha\beta,(1)}_n, \\
  \sigma^{\alpha\beta,(0)}_n&=& \frac{-e^2}{4\pi^3}\int\!d^2k\, v_{k,n}^\alpha v_{k,n}^\beta
  \tau_{k,n} \frac{\partial f_k^0}{\partial\varepsilon_{k,n}}, \\
  \sigma^{\alpha\beta,(1)}_n &=& \frac{e^3}{4\pi^3\hbar}\int\!d^2k\,
  \frac{\partial f_k^0}{\partial\varepsilon_{k,n}} v_{k,n}^\alpha \tau_{k,n} \   \left(\vec{v}_{k,n}
  \times \vec{B}\right) \cdot \vec{\nabla}_k\left( v_{k,n}^\beta\tau_{k,n}\right) \,,
\end{eqnarray*}
where $\alpha,\beta=x,y$, $e$ is the electron charge, $ v_{k,n}^\alpha$ is the $\alpha$ component of the group velocity of a Bloch electron with wavevector $k$ in the $n$-th band, $\tau_{k,n}$ is the scattering time for a Bloch electron with wavevector $k$ in the $n$-th band, and $f_k^0=[\mathrm e^{(\varepsilon_{k,n}-\mu)/T}+1]^{-1}$ is the Fermi distribution function for a Bloch electron with energy $\varepsilon_{k,n}$ at a temperature $T$, $\mu$ being the chemical potential. Notice that also for a nematic system $\sigma_n^{\alpha\ne\beta,(0)}=0$, so that only the diagonal elements of $\sigma^{\alpha\beta,(0)}_n$ contribute. Moreover, for a two-dimensional system in the $xy$-plane only a magnetic field along the $z$-direction contributes to $\sigma^{\alpha\beta,(1)}_n$, which then only contains non-diagonal elements. Finally, the Onsager reciprocity requires $\hat{\sigma}_n(B_z)=\hat{\sigma}_n^\mathrm{T}(-B_z)$, where $\mathrm T$ implies taking the transpose matrix.

The inverse of Eq.\,(\ref{eq:sigma}) is $\vec{E}=\hat{\rho}_n\vec{J}_n$ with
\begin{eqnarray}
  \hat{\rho}_n=\hat{\sigma}_n^{-1}=\frac{1}{\sigma_n^{xx\phantom y}\sigma_n^{yy}+(\sigma_n^{xy})^2}\left(
  \begin{array}{cc}
    \sigma_n^{yy} & -\sigma^{xy}_n \\
    \sigma^{xy}_n & \sigma_n^{xx}
  \end{array}\right) 
  \equiv \left(
  \begin{array}{cc}
    \rho_n^{x} & B_z R_n^H \\
    -B_z R_n^H & \rho_n^{y}
  \end{array}\right)\,.
\end{eqnarray}
For a $N-$band system we can now generalize this expression to $\vec{E}=\sum_{n=1}^N \hat{\rho}_n\vec{J}_n \equiv \hat{\rho} \vec{J} $ where $\vec{J}=\sum_n \vec{J}_n$. The expression for the full resistivity tensor is given by
\begin{eqnarray}
  \hat{\rho}^{-1}&=&\sum_{n=1}^{N} \hat{\rho}_n^{-1}=\sum_{n=1}^{N} \hat{\sigma}_n 
  = \sum_{n=1}^{N}
\frac{1}{\rho_n^{x\phantom y} \rho_n^y +(B_z R_n^H)^2}  \left(
  \begin{array}{cc}
    \rho_n^y & -B_z R_n^H \\
    B_z R_n^H & \rho_n^x
  \end{array}\right) \nonumber \,,
\end{eqnarray}
which, upon defining $\sigma_n^{x,y}\equiv 1/\rho_n^{x,y}$, can be written as
\begin{equation}
  \hat{\rho}^{-1}= \sum_{n=1}^{N}
\frac{1}{1 +\sigma_n^x\sigma_n^y (B_z R_n^H)^2}
  \left(
  \begin{array}{cc}
    \sigma_n^x & -\sigma_n^x\sigma_n^y B_z R_n^H \\
    \sigma_n^x\sigma_n^y B_z R_n^H & \sigma_n^y
  \end{array}\right)
  \equiv
  \left(
  \begin{array}{cc}
    \rho^x & B_z R^H \\
    -B_z R^H & \rho^y
  \end{array}\right)^{-1}
  \,,
\end{equation}
that is the general definition of the resistivity in a multiband system used and discussed in the main text. For $N=2$ bands the explicit expressions read
\begin{eqnarray}
    \rho^x &=& \frac{\sigma_1^y+\sigma_2^y +\sigma_1^y\sigma_2^y\left\lbrack\sigma_1^x (R_1^H)^2+\sigma_2^x (R_2^H)^2
    \right\rbrack B_z^2}{\sigma_1^x\sigma_1^y+\sigma_2^x\sigma_2^y+\sigma_1^x\sigma_2^y+\sigma_2^x\sigma_1^y+
    \sigma_1^x\sigma_2^x\sigma_1^y\sigma_2^y(R_1^H+R_2^H)^2 B_z^2}\,, \label{eq:rx}\\
   && \nonumber \\
  \rho^y &=& \frac{\sigma_1^x+\sigma_2^x +\sigma_1^x\sigma_2^x\left\lbrack\sigma_1^y (R_1^H)^2+\sigma_2^y (R_2^H)^2
  \right\rbrack B_z^2}{\sigma_1^x\sigma_1^y+\sigma_2^x\sigma_2^y+\sigma_1^x\sigma_2^y+\sigma_2^x\sigma_1^y+
  \sigma_1^x\sigma_2^x\sigma_1^y\sigma_2^y(R_1^H+R_2^H)^2 B_z^2}\,, \label{eq:ry}\\
  && \nonumber \\
    R^H &=& \frac{\sigma_1^x\sigma_1^y R_1^H + \sigma_2^x\sigma_2^y R_2^H+\sigma_1^x\sigma_1^y\sigma_2^x\sigma_2^y R_1^H R_2^H 
    ( R_1^H + R_2^H)B_z^2 }{\sigma_1^x\sigma_1^y+\sigma_2^x\sigma_2^y+\sigma_1^x\sigma_2^y+\sigma_2^x\sigma_1^y+
    \sigma_1^x\sigma_2^x\sigma_1^y\sigma_2^y(R_1^H+R_2^H)^2 B_z^2}\,,\label{eq:rb}
  \end{eqnarray}
which, in the isotropic case ($\sigma_1^x=\sigma_1^y\equiv \sigma_1$, $\sigma_2^x=\sigma_2^y\equiv \sigma_2$), reduce to the 
well-known results
\begin{eqnarray}
\rho^x=\rho^y &=& \frac{\sigma_1+\sigma_2+\sigma_1\sigma_2\left\lbrack \sigma_1 (R_1^H)^2 + \sigma_2(R_2^H)^2
\right\rbrack B_z^2}{(\sigma_1+\sigma_2)^2 +\sigma_1^2\sigma_2^2 (R_1^H+R_2^H)^2 B_z^2}\,, \label{eq:r2x}\\
&& \nonumber \\
    R^H &=& \frac{\sigma_1^2 R_1^H + \sigma_2^2 R_2^H+\sigma_1^2\sigma_2^2 R_1^H R_2^H 
    ( R_1^H + R_2^H)B_z^2 }{(\sigma_1+\sigma_2)^2 +\sigma_1^2\sigma_2^2 (R_1^H+R_2^H)^2 B_z^2}\,. \label{eq:r2b}
\end{eqnarray}
In terms of the individual resistivities $\rho_{1,2}^{x,y}$, Eqs.\,(\ref{eq:rx},\ref{eq:ry},\ref{eq:rb}) can be rewritten as
\begin{eqnarray}
    \rho^x &=& \frac{\rho_1^x\rho_2^x + (\rho_1^y+\rho_2^y)^{-1}\left\lbrack\rho_2^x (R_1^H)^2+\rho_1^x (R_2^H)^2
    \right\rbrack B_z^2}{\rho_1^x+\rho_2^x+(\rho_1^y+\rho_2^y)^{-1}(R_1^H+R_2^H)^2 B_z^2}, \label{eq:rx2}\\
    && \nonumber \\
  \rho^y &=& \frac{\rho_1^y\rho_2^y  +(\rho_1^y+\rho_2^y)^{-1}\left\lbrack\rho_2^y (R_1^H)^2+\rho_1^y (R_2^H)^2
  \right\rbrack B_z^2}{\rho_1^y+\rho_2^y+(\rho_1^y+\rho_2^y)^{-1}(R_1^H+R_2^H)^2 B_z^2}\,, \label{eq:ry2}\\
  && \nonumber \\
    R^H &=& \frac{\rho_2^x\rho_2^y R_1^H + \rho_1^x\rho_1^y R_2^H+R_1^H R_2^H 
    ( R_1^H + R_2^H)B_z^2 }{\rho_1^x\rho_1^y+\rho_2^x\rho_2^y+\rho_1^x\rho_2^y+\rho_2^x\rho_1^y+(R_1^H+R_2^H)^2 B_z^2}\,.\label{eq:rb2}
  \end{eqnarray}

\subsection{Hall coefficient}
In both the isotropic and anisotropic case, the Hall coefficient can be written in the form
\begin{equation}\label{eq:fit_RH}
  R^H=\frac{a_0 + a_1 B_z^2}{1+a_2 B_z^2}\,,
\end{equation}
where, in the isotropic case the coefficients are given by
\begin{eqnarray}
  a_0&=& \frac{\sigma_1^2 R_1^H + \sigma_2^2 R_2^H}{(\sigma_1+\sigma_2)^2}\,, \\
  && \nonumber \\
  a_1&=& \left(\frac{\sigma_1\sigma_2}{\sigma_1+\sigma_2}\right)^2 R_1^H R_2^H
  (R_1^H+R_2^H)\,, \\
  && \nonumber \\
  a_2&=& \left(\frac{\sigma_1\sigma_2}{\sigma_1+\sigma_2}\right)^2 (R_1^H+R_2^H)^2 \,,
  \end{eqnarray}
whereas in the anisotropic case they read
\begin{eqnarray}
  a_0&=&\frac{\sigma_1^x\sigma_1^y R_1^H + \sigma_2^x\sigma_2^y R_2^H}{\sigma_1^x\sigma_1^y+\sigma_2^x\sigma_2^y+\sigma_1^x\sigma_2^y+\sigma_2^x\sigma_1^y}\,,\\
  && \nonumber \\
  a_1&=& \frac{\sigma_1^x\sigma_1^y\sigma_2^x\sigma_2^y R_1^H R_2^H ( R_1^H + R_2^H)}{\sigma_1^x\sigma_1^y+\sigma_2^x\sigma_2^y+\sigma_1^x\sigma_2^y+\sigma_2^x\sigma_1^y}\,, \\
  && \nonumber \\
  a_2 &=& \frac{\sigma_1^x\sigma_2^x\sigma_1^y\sigma_2^y(R_1^H+R_2^H)^2}{\sigma_1^x\sigma_1^y+\sigma_2^x\sigma_2^y+\sigma_1^x\sigma_2^y+\sigma_2^x\sigma_1^y}\,.
\end{eqnarray}

\subsection{Magnetoresistance}
Similarly, in both the isotropic and anisotropic case, the MR can be written as
\begin{equation}\label{eq:fit_magneto}
\frac{\Delta \rho^x(B)}{\rho^x(0)} =  \frac{\rho^x(B)}{\rho^x(0)}-1=\frac{1+b_0 B_z^2}{1+b_1 B_z^2}-1.
\end{equation}
In the isotropic case the coefficients read
\begin{eqnarray}
  b_0&=&\frac{\sigma_1\sigma_2}{\sigma_1+\sigma_2}\left\lbrack \sigma_1 (R_1^H)^2 + \sigma_2 (R_2^H)^2 \right\rbrack\,, 
  \label{eq:c0}\\
  && \nonumber \\
  b_1 &=& \left(\frac{\sigma_1\sigma_2}{\sigma_1+\sigma_2}\right)^2
  \left(R_1^H + R_2^H\right)^2\,,
\end{eqnarray}  
whereas in the anisotropic case they are given by
\begin{eqnarray}
 b_0&=& \frac{\sigma_1^y\sigma_2^y\left\lbrack\sigma_1^x (R_1^H)^2+\sigma_2^x (R_2^H)^2
 \right\rbrack}{\sigma_1^y+\sigma_2^y}\,,\label{eq:c02}\\
&&  \nonumber \\
b_1&=& \frac{\sigma_1^x\sigma_2^x\sigma_1^y\sigma_2^y(R_1^H+R_2^H)^2}{\sigma_1^x\sigma_1^y+\sigma_2^x\sigma_2^y+\sigma_1^x\sigma_2^y+\sigma_2^x\sigma_1^y} \,.
\end{eqnarray}
Note that in the latter case the coefficient $b_0$ is derived from $\rho^x(B_z)$, i.e., assuming that the electric field is applied along the $x$-direction. When the electric field is applied along the $y$-direction, the replacement $x (y)\to x (y)$ must be implemented in Eq.\,(\ref{eq:c02}).

\subsection{Magneto-elasto-resistance}
We define the total resistivity anisotropy as
$N = 2 (\rho^x-\rho^y)/(\rho^x+\rho^y)$ and similarly for the
individual components $N_n = 2 (\rho_n^x-\rho_n^y)/(\rho_n^x+\rho_n^y)$.
Then, from Eqs.\,(\ref{eq:rx},\ref{eq:ry}), one obtains
\begin{equation}
  N=2\frac{\left\lbrack \rho_2^x \rho_2^y+(R_2^H)^2 B_z^2\right\rbrack
    (\rho_1^x-\rho_1^y)
    + \left\lbrack \rho_1^x\rho_1^y+(R_1^H)^2 B_z^2\right\rbrack
    (\rho_2^x-\rho_2^y)}
  {\left\lbrack \rho_2^x\rho_2^y+(R_2^H)^2 B_z^2 \right\rbrack
    (\rho_1^x+\rho_1^y)
    + \left\lbrack \rho_1^x\rho_1^y+(R_1^H)^2 B_z^2 \right\rbrack
    (\rho_2^x+\rho_2^y)}\,.
\end{equation}

\noindent Assuming that, in the simplest case,
\begin{equation}
  \rho_n^{x/y}=\rho_n^{x/y,0} \pm \frac{1}{2}\gamma_n \epsilon ,
  \label{rho_epsilon}
\end{equation}
and defining
\begin{eqnarray}
  \rho_n^0 = \frac{1}{2}\left\lbrack \rho_n^{x,0}+\rho_n^{y,0}\right\rbrack, \quad \quad \quad
  N_n^0 =  \frac{\rho_n^{x,0}-\rho_n^{y,0}}{\rho_n^0}, \quad \quad \quad
    M_n^0 &=& \frac{\rho_n^{x,0}\rho_n^{y,0}+(R_i^H)^2 B^2}{(\rho^0_n)^2} ,
\end{eqnarray}
one then finds  
\begin{eqnarray}\label{eq:elres}
  \frac{\partial N}{\partial \epsilon}&=&\frac{1}{\rho_1^0 M_1^0+\rho_2^0 M_2^0}
  \left( \gamma_1{\frac{\rho_2^0}{\rho_1^0}}M_2^0
  +\gamma_2 \frac{\rho_1^0}{\rho_2^0} M_1^0\right)\nonumber\\
  &+&\frac{1}{2}\frac{N_1^0-N_2^0}{(\rho_1^0 M_1^0+\rho_2^0 M_2^0)^2}
  \left( \gamma_1 N_1^0 \rho_2^0 M_2^0-\gamma_2 N_2^0 \rho_1^0 M_1^0  \right)\,.
\end{eqnarray}

Notice that the whole dependence of the MER on the magnetic field is contained in $M^0_n$, where $B$ appears 
multiplied by the $n$-th band Hall coefficient $R_n$. As a consequence, one can immediately see that for in-plane magnetic 
field $R_n=0$, so that the MER does not depend anymore on $B$, in agreement with our experimental findings. 

In the isotropic case, using $N_n^0=0$, which implies
$M_n=1+R_n^2B^2/(\rho_n^0)^2$, Eq.\,(\ref{eq:elres}) reduces to
\begin{equation}
\label{eq:ER_general}
\frac{\partial N}{\partial \epsilon}=\frac{\gamma_1\frac{\rho_2^0}{\rho_1^0}M_2^0+\gamma_2
\frac{\rho_1^0}{\rho_2^0}M_1^0}{\rho_1^0 M_1^0+\rho_2^0 M_2^0}=
\frac{\gamma_1^0\frac{\rho_2^0}{\rho_1^0}+\gamma_2\frac{\rho_1^0}{\rho_2^0}+
\frac{\gamma_1(R_2^H)^2+\gamma_2(R_1^H)^2}{\rho_1^0\rho_2^0}B_z^2}{\rho_1^0+\rho_2^0+\left(\frac{(R_1^H)^2}{\rho_1^0}+
\frac{(R_2^H)^2}{\rho_2^0}\right)B_z^2}\approx \frac{\gamma}{\rho}\,,
\end{equation}
where the last step involves the approximation $\rho_n^0\equiv \rho$, $\gamma_n \equiv \gamma$ and $R_n^H \equiv R_H$. 
In this limit the MER is completely independent of magnetic field. Notice however that, since above $T_S$ the Hall coefficient $R_H$ is of order $0.1\,\mu\Omega/$T and the resistivity $\rho$ is of order $100\,\mu\Omega)$, the MER magnetic field dependence would be quite weak even without the above approximations, in agreement with the experimental result.
Finally, we note that in the present model the temperature dependence of the coefficient $\gamma\sim 1/(T-T_S)$ has to be assumed phenomenologically, but could be in principle derived via a coupled Ginzburg-Landau theory.

We conclude deriving expressions suitable for the fit procedure. Let us define
\begin{eqnarray*}
\rho_1^0 M_1^0+\rho_2^0 M_2^0  &\equiv&  K_1+ K_2 B_z^2\,,\\
&& \nonumber \\
\gamma_1\frac{\rho_2^0}{\rho_1^0} M_2^0+\gamma_2\frac{\rho_1^0}{\rho_2^0} M_1^0 &\equiv& K_3+K_4 B_z^2\,, \\ 
&& \nonumber \\
\gamma_1 N_1^0 \rho_2^0 M_2^0-\gamma_2 N_2^0 \rho_1^0 M_1^0 &\equiv& K_5+K_6 B_z^2\,,
\end{eqnarray*}
where 
\begin{eqnarray*}
  K_1 &=&\rho_1^0 \left[ 1-\frac{1}{4}(N_1^0)^2\right]
  +\rho_2^0\left[ 1-\frac{1}{4}(N_2^0)^2\right]\,, 
  \quad \quad \quad \quad \quad \ \
  K_2 =  \left(\frac{(R_1^H)^2}{\rho_1^0}+\frac{(R_2^H)^2}{\rho_2^0}\right),\\
  K_3 &=&\gamma_1\frac{\rho_2^0}{\rho_1^0} \left[ 1-\frac{1}{4}(N_2^0)^2\right] +\gamma_2\frac{\rho_1^0}{\rho_2^0}
  \left[ 1-\frac{1}{4}(N_1^0)^2\right]\,,
  \quad \quad \quad 
  K_4 = \frac{\gamma_1 (R_2^H)^2+\gamma_2 (R_1^H)^2}{\rho_1^0\rho_2^0},\\ 
  K_5 &=& \gamma_1 N_1^0 \rho_2^0 \left[ 1-\frac{1}{4}(N_2^0)^2\right]
  -\gamma_2 N_2^0 \rho_1^0 \left[ 1-\frac{1}{4}(N_1^0)^2\right]\,, 
  \quad K_6 = \left( \frac{\gamma_1 N_1^0  (R_2^H)^2}{\rho_2^0} 
  -  \frac{\gamma_2 N_2^0(R_1^H)^2}{\rho_1^0}\right)\,,
\end{eqnarray*}
so that Eq.\,(\ref{eq:elres}) reads
\begin{equation}
  \frac{\partial N}{\partial \epsilon}=\frac{K_3+K_4B_z^2}{K_1 + K_2 B_z^2}
  +\frac{N_1^0-N_2^0}{2}\frac{K_5+K_6B_z^2}{(K_1+K_2B_z^2)^2}\,.
\end{equation}
The fitting formula then reads
\begin{equation}\label{eq:fit_elastofull}
  \frac{\partial N}{\partial \epsilon}=c_0\frac{1+c_1 B_z^2}{1 + c_2  B_z^2}
  +c_3 \frac{1 + c_4 B_z^2}{(1+c_2 B_z^2)^2}\,,
\end{equation}
with
\begin{eqnarray*}
  c_0&=&\frac{K_3}{K_1}=\frac{\gamma_1\frac{\rho_2^0}{\rho_1^0} \left[ 1-\frac{1}{4}(N_2^0)^2\right]
  +\gamma_2\frac{\rho_1^0}{\rho_2^0}\left[ 1-\frac{1}{4}(N_1^0)^2\right]}{\rho_1^0 \left[ 1-\frac{1}{4}(N_1^0)^2\right]
  +\rho_2^0\left[ 1-\frac{1}{4}(N_2^0)^2\right]}\,,\\
    &&\nonumber \\
  c_1&=&\frac{K_4}{K_3}=\frac{\frac{\gamma_1 (R_2^H)^2+\gamma_2 (R_1^H)^2}{\rho_1^0\rho_2^0}}{\gamma_1\frac{\rho_2^0}{\rho_1^0} 
  \left[ 1-\frac{1}{4}(N_2^0)^2\right]
  +\gamma_2\frac{\rho_1^0}{\rho_2^0}\left[ 1-\frac{1}{4}(N_1^0)^2\right]}\,,\\
    &&\nonumber \\
  c_2&=&\frac{K_2}{K_1}=\frac{\frac{(R_1^H)^2}{\rho_1^0}+\frac{(R_2^H)^2}{\rho_2^0}}{\rho_1^0 \left[ 1-\frac{1}{4}(N_1^0)^2\right]
  +\rho_2^0\left[ 1-\frac{1}{4}(N_2^0)^2\right]}\,,\\
  &&\nonumber \\
c_3&=&\frac{N_1^0-N_2^0}{2} \frac{K_5}{K_1^2}=\frac{N_1^0-N_2^0}{2} \frac{\gamma_1 N_1^0 \rho_2^0 \left[ 1-\frac{1}{4}(N_2^0)^2\right]
  -\gamma_2 N_2^0 \rho_1^0 \left[ 1-\frac{1}{4}(N_1^0)^2\right]}{\left\{\rho_1^0 \left[ 1-\frac{1}{4}(N_1^0)^2\right]
  +\rho_2^0\left[ 1-\frac{1}{4}(N_2^0)^2\right] \right\}^2}\,,\\
    &&\nonumber \\
c_4&=&\frac{K_6}{K_5}=\frac{\left( \gamma_1 N_1^0 \frac{(R_2^H)^2}{\rho_2^0}-\gamma_2 N_2^0 \frac{(R_1^H)^2}{\rho_1^0}\right)}{\gamma_1 N_1^0 \rho_2^0 \left[ 1-\frac{1}{4}(N_2^0)^2\right]
  -\gamma_2 N_2^0 \rho_1^0 \left[ 1-\frac{1}{4}(N_1^0)^2\right]}\,.
\end{eqnarray*}
Notice that, if we set $N_n^0=0$, the expression reduces to 
\begin{equation}\label{eq:fit_elasto}
  \frac{\partial N}{\partial \epsilon}=c_0\frac{1+c_1 B_z^2}{1 + c_2 B_z^2}\,,
\end{equation}
with fitting parameters given by 
\begin{eqnarray}
c_0&=&\frac{1}{\rho_1^0+\rho_2^0}\frac{1}{\rho_1^0\rho_2^0}\left[\gamma_1(\rho_2^0)^2+\gamma_2(\rho_1^0)^2 \right]\,,\\
&& \nonumber \\ 
c_1&=&\frac{\gamma_1 (R_2^H)^2+\gamma_2 (R_1^H)^2}{\gamma_1(\rho_2^0)^2+\gamma_2(\rho_1^0)^2}\,,\\
&& \nonumber \\
c_2&=&\frac{1}{\rho_1^0 +\rho_2^0}\left(\frac{(R_1^H)^2}{\rho_1^0}+\frac{(R_2^H)^2} {\rho_2^0}\right)\,.
\end{eqnarray}
The assumption $N_n^0=0$ means that $\rho_n^{x,0}-\rho_n^{y,0} =0$. This is certainly true in the isotropic case, however, it can also be assumed in the nematic phase if we imply that the anisotropy is proportional to the strain only, see Eq.\,(\ref{rho_epsilon}).

\section{Fit of the experimental results}
\label{App:Fit}

In this appendix we summarize the results of the fit procedure for Hall resistivity, MR and MER, using Eq.s\,(\ref{eq:fit_RH}, \ref{eq:fit_magneto}, \ref{eq:fit_elastofull}). 

In the main text, we show in Fig\,\ref{Figure2}(b) the fit of the experimental data for the Hall resistivity using Eq.\,(\ref{eq:fit_RH}). As one can see, the procedure yields a rather good fit to the data for $\rho_{xy}= R^H B_z$. The fitting parameters are given in Tab.\,\ref{tab1}.

\begin{table}
\begin{equation}
\begin{array}{|c|c|c|c|}
  T \,[\mathrm K] & a_0\, [\mu\Omega\,\mathrm{cm/T}] & a_1 \,[\mu\Omega\,\mathrm{cm/T}^3] & a_2 \, [1/\mathrm{T}^2] \\
  \hline
  15 & -0.3931 & 0.0029 & 0.02571 \\
  20 & -0.466 & 0.00135 & 0.0271 \\
  30 & -0.3191 & 0.00159 & 0.01638 \\
  40 & -0.1239 & 0.00163 & 0.00976 \\
  50 & -0.057 & 0.00086 & 0.00868 \\
  60 & 0.02423 & 0.000456 & 0.003967 \\
  80 & 0.0823 & -1.265\times 10^{-5} & 7.698\times 10^{-5} \\
  90 & 0.0556 & -1.51\times 10^{-5} & 1.874\times 10^{-4} \\
  120 & 0.0269 & -0.00216 & 0.30396
\end{array}
\end{equation}
\caption{Parameters for fitting the Hall coefficient Eq.\,(\ref{eq:fit_RH})
  to experimental data.}
\label{tab1}
\end{table}
\begin{table}
\begin{equation}
\begin{array}{|c|c|c|}
  T \,[\mathrm K] & b_0\,[1/\mathrm{T}^2]  & b_1\,[1/\mathrm{T}^2] \\
  \hline
  15 & 0.01078 & 0.00364 \\
  20 & 0.00922 & 0.003495 \\
  30 & 0.00673 & 0.00333 \\
  40 & 0.00636 & 0.00407 \\
  50 & 0.00346 & 0.00231 \\
  60 & 0.01195 & 0.01063 \\
  90 & 0.28413 & 0.27859 
\end{array}
\end{equation}
\caption{Parameters for fitting the MR Eq.\,(\ref{eq:fit_magneto}) to experimental data.}
\label{tab2}
\end{table}
\begin{table}
\begin{equation}
\begin{array}{|c|c|c|c|c|}
  T \,[\mathrm K] & a_0\, [\mu\Omega\,\mathrm{cm/T}] & a_1 \,[\mu\Omega\,\mathrm{cm/T}^3] & b_0\, [1/\mathrm{T}^2]& b_1=a_2 \, [1/\mathrm{T}^2] \\
  \hline
  15 &  -0.277695   &     0.00197007  &   0.0215466    &   0.01 \\
  20 &  -0.340067   &     0.00159451  &   0.0161647    &   0.008 \\
  30 &  -0.262895   &     0.00139701  &   0.013011     &   0.008 \\
  40 &  -0.110791   &     0.00140892  &   0.00992253   &   0.007 \\
  50 &  -0.0463063  &     0.00066924  &   0.00650638   &   0.005 \\
  60 & 0.020013   &     0.00066443  &   0.00804311   &   0.007 \\
  90 & 0.0653995  &     0.00769684  &   0.153141     &   0.15
\end{array}
\end{equation}
\caption{Parameters for fitting the Hall coefficient Eq.\,(\ref{eq:fit_RH}) the MR Eq.\,(\ref{eq:fit_magneto}) to experimental 
data, with the constraint $b_1=a_2$.}
\label{tab2-bis}
\end{table}

\begin{figure}[hhh!]
\includegraphics[angle=0,scale=0.4]{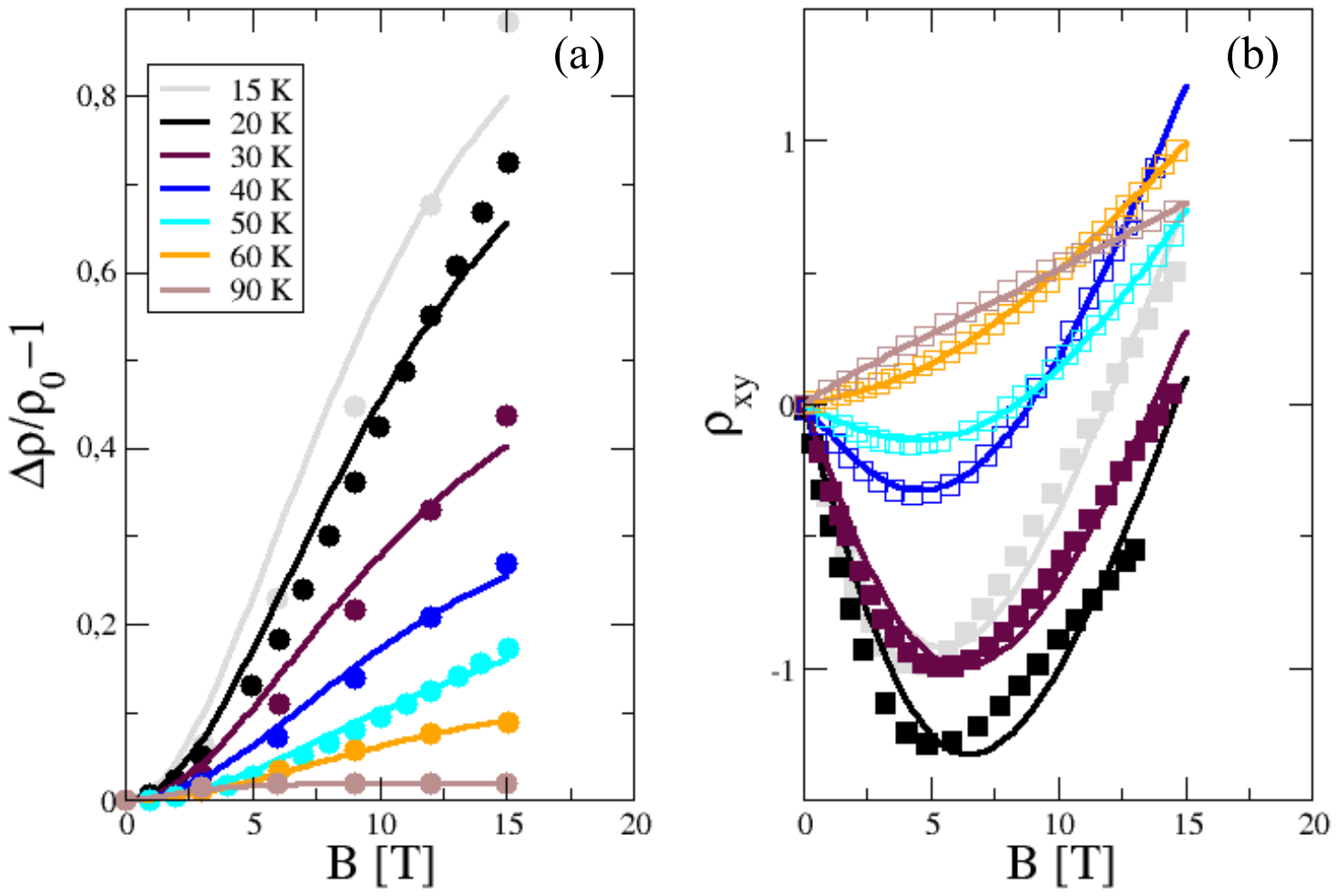}\caption{Simultaneous fit of the MR and of the Hall resistivity, with the constraint $b_1=a_2$. In both panels (a) and (b), the theoretical fit using Eq.s\,(\ref{eq:magneto}) and (\ref{eq:RH}) are shown as solid lines on top of the experimental data (symbols).}
\label{fig-constraint}
\end{figure}

The analysis of the MR is shown in Fig.\,\ref{Figure2}(a), where the theoretical fit is shown on top of the experimental data using Eq\,(\ref{eq:fit_magneto}). The corresponding fitting parameters are listed in Tab.\,\ref{tab2}.
As we already remarked in the main text, within the present derivation, the parameters $a_2$ and $b_1$ describe the same parameter combination. However, the fit procedure lead to a difference of approximately a factor of $7-8$ at low temperature. In principle, one can simultaneously fit the Hall resistivity and the MR, constraining $b_1=a_2$, finding 
the results shown in Tab.\,\ref{tab2-bis}. The overall trend of the fitting parameters is maintained, although the introduction of a constraint worsens the fit, especially at low temperature, as it can be seen in Fig.\,\ref{fig-constraint}. However, it must be pointed out that the parameters $b_1$ and $a_2$ are only relevant for fitting the MR and $\rho^{xy}$ at large fields. The discrepancy in the unconstrained fit may then point toward a violation of the standard perturbative treatment of the Boltzmann equation, as discussed in the main text.
 
\begin{figure}[hhh!]
\includegraphics[angle=-0,scale=0.4]{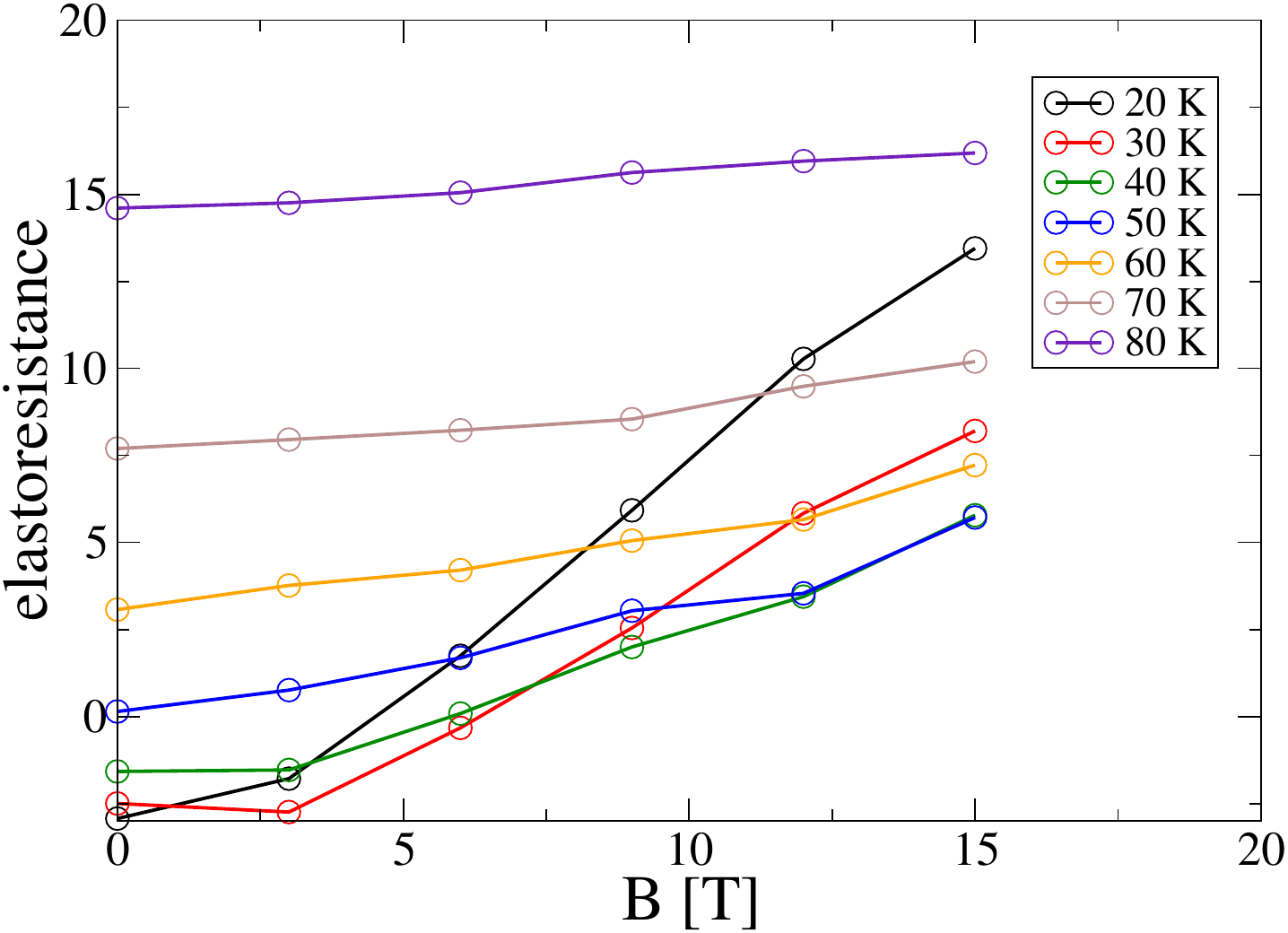}
\caption{Magnetic-field dependence of the MER extracted form the polynomial fits at various temperatures below $T_S$.}
\label{fig:6}
\end{figure}

The analysis of the MER is shown in Fig.\,\ref{Figure3}, where we show the experimental data together with the theoretical polynomial fits Eq.\,(\ref{eq:fit_elastofull}). From these fits we have extracted the magnetic field dependence at different temperatures. The corresponding curves are presented in Fig.\,\ref{fig:6}, which shows a positive trend for all temperatures and an enhanced field-dependence in the temperature range above $T_C$ where superconducting fluctuations are expected to be relevant.

Upon fitting the MER using Eq.\,(\ref{eq:fit_elastofull}), one realizes that the inclusion of coefficients $c_3$ and $c_4$ does not improve the fit. Alternatively, one could also do the fitting with only coefficients $c_2$, $c_3$, $c_4$ with a similar accuracy, however, the first choice implies that $N_n^0 \approx 0$ whereas the second choice is harder to justify. Let us then assume $N_n^0 \approx 0$ so that the MER can be written as Eq.\,(\ref{eq:fit_elasto}). In this approximation, we find the set shown in Tab.\,\ref{tab3}. 

\begin{table}
\begin{equation}
\begin{array}{|c|c|c|c|}
  T \,[\mathrm K]  & c_0  & c_1  \, [1/\mathrm{T}^2] & c_2 \, [1/\mathrm{T}^2]\\
  \hline
  20  & -3.037 & -0.0467 & 0.005\\
  25  & -3.2 & -0.032 & 0.004 \\
  30  & -2.991 & -0.0255 & 0.003   \\
  35  & -2.476 & -0.024 & 0.0028 \\
  40  & -1.724 & -0.0028 & 0.0026  \\
  45  & -0.775 & -0.052 & 0.0025 \\
  50  & 0.367 & 0.096  & 0.0023\\
  55  & 1.73 & 0.0177  & 0.0018\\
  60  & 3.36 & 0.0073 & 0.0011\\
  65  & 5.33 & 0.0034 & 0.0005 \\
  70  & 7.76 & 0.0018  & 0.0003 \\
  75  & 10.8 & 0.002 & 0.0011 \\
  80  & 14.5821 & 0.0087 & 0.00742
\end{array}
\end{equation}
\caption{Parameters for fitting the MER Eq.\,(\ref{eq:fit_elasto}) to experimental data.}
\label{tab3}
\end{table}


\clearpage


\begin{thebibliography}{40}%
\makeatletter
\providecommand \@ifxundefined [1]{%
 \@ifx{#1\undefined}
}%
\providecommand \@ifnum [1]{%
 \ifnum #1\expandafter \@firstoftwo
 \else \expandafter \@secondoftwo
 \fi
}%
\providecommand \@ifx [1]{%
 \ifx #1\expandafter \@firstoftwo
 \else \expandafter \@secondoftwo
 \fi
}%
\providecommand \natexlab [1]{#1}%
\providecommand \enquote  [1]{``#1''}%
\providecommand \bibnamefont  [1]{#1}%
\providecommand \bibfnamefont [1]{#1}%
\providecommand \citenamefont [1]{#1}%
\providecommand \href@noop [0]{\@secondoftwo}%
\providecommand \href [0]{\begingroup \@sanitize@url \@href}%
\providecommand \@href[1]{\@@startlink{#1}\@@href}%
\providecommand \@@href[1]{\endgroup#1\@@endlink}%
\providecommand \@sanitize@url [0]{\catcode `\\12\catcode `\$12\catcode `\&12\catcode `\#12\catcode `\^12\catcode `\_12\catcode `\%12\relax}%
\providecommand \@@startlink[1]{}%
\providecommand \@@endlink[0]{}%
\providecommand \url  [0]{\begingroup\@sanitize@url \@url }%
\providecommand \@url [1]{\endgroup\@href {#1}{\urlprefix }}%
\providecommand \urlprefix  [0]{URL }%
\providecommand \Eprint [0]{\href }%
\providecommand \doibase [0]{https://doi.org/}%
\providecommand \selectlanguage [0]{\@gobble}%
\providecommand \bibinfo  [0]{\@secondoftwo}%
\providecommand \bibfield  [0]{\@secondoftwo}%
\providecommand \translation [1]{[#1]}%
\providecommand \BibitemOpen [0]{}%
\providecommand \bibitemStop [0]{}%
\providecommand \bibitemNoStop [0]{.\EOS\space}%
\providecommand \EOS [0]{\spacefactor3000\relax}%
\providecommand \BibitemShut  [1]{\csname bibitem#1\endcsname}%
\let\auto@bib@innerbib\@empty
\bibitem [{\citenamefont {Fernandes}\ \emph {et~al.}(2022)\citenamefont {Fernandes}, \citenamefont {Coldea}, \citenamefont {Ding}, \citenamefont {Fisher}, \citenamefont {Hirschfeld},\ and\ \citenamefont {Kotliar}}]{Nature2022}%
  \BibitemOpen
  \bibfield  {author} {\bibinfo {author} {\bibfnamefont {R.~M.}\ \bibnamefont {Fernandes}}, \bibinfo {author} {\bibfnamefont {A.~I.}\ \bibnamefont {Coldea}}, \bibinfo {author} {\bibfnamefont {H.}~\bibnamefont {Ding}}, \bibinfo {author} {\bibfnamefont {I.~R.}\ \bibnamefont {Fisher}}, \bibinfo {author} {\bibfnamefont {P.~J.}\ \bibnamefont {Hirschfeld}},\ and\ \bibinfo {author} {\bibfnamefont {G.}~\bibnamefont {Kotliar}},\ }\bibfield  {title} {\bibinfo {title} {Iron pnictides and chalcogenides: a new paradigm for superconductivity},\ }\href {https://doi.org/10.1038/s41586-021-04073-2} {\bibfield  {journal} {\bibinfo  {journal} {Nature}\ }\textbf {\bibinfo {volume} {601}},\ \bibinfo {pages} {35} (\bibinfo {year} {2022})}\BibitemShut {NoStop}%
\bibitem [{\citenamefont {Fernandes}\ \emph {et~al.}(2014)\citenamefont {Fernandes}, \citenamefont {Chubukov},\ and\ \citenamefont {Schmalian}}]{NatPhys2014}%
  \BibitemOpen
  \bibfield  {author} {\bibinfo {author} {\bibfnamefont {R.~M.}\ \bibnamefont {Fernandes}}, \bibinfo {author} {\bibfnamefont {A.}~\bibnamefont {Chubukov}},\ and\ \bibinfo {author} {\bibfnamefont {J.}~\bibnamefont {Schmalian}},\ }\bibfield  {title} {\bibinfo {title} {What drives nematic order in iron-based superconductors?},\ }\href {https://doi.org/10.1038/nphys2877} {\bibfield  {journal} {\bibinfo  {journal} {Nature Physics}\ }\textbf {\bibinfo {volume} {10}},\ \bibinfo {pages} {97} (\bibinfo {year} {2014})}\BibitemShut {NoStop}%
\bibitem [{\citenamefont {Wang}\ \emph {et~al.}(2022)\citenamefont {Wang}, \citenamefont {Fanfarillo},\ and\ \citenamefont {Böhmer}}]{Qisi_Frontiers2022}%
  \BibitemOpen
  \bibfield  {author} {\bibinfo {author} {\bibfnamefont {Q.}~\bibnamefont {Wang}}, \bibinfo {author} {\bibfnamefont {L.}~\bibnamefont {Fanfarillo}},\ and\ \bibinfo {author} {\bibfnamefont {A.~E.}\ \bibnamefont {Böhmer}},\ }\bibfield  {title} {\bibinfo {title} {Nematicity in iron-based superconductors},\ }\bibfield  {journal} {\bibinfo  {journal} {Frontiers in Physics}\ }\textbf {\bibinfo {volume} {10}},\ \href {https://doi.org/10.3389/fphy.2022.1038127} {10.3389/fphy.2022.1038127} (\bibinfo {year} {2022})\BibitemShut {NoStop}%
\bibitem [{\citenamefont {B{\"o}hmer}\ \emph {et~al.}(2022)\citenamefont {B{\"o}hmer}, \citenamefont {Chu}, \citenamefont {Lederer},\ and\ \citenamefont {Yi}}]{Bohmer_NatPhys2022}%
  \BibitemOpen
  \bibfield  {author} {\bibinfo {author} {\bibfnamefont {A.~E.}\ \bibnamefont {B{\"o}hmer}}, \bibinfo {author} {\bibfnamefont {J.-H.}\ \bibnamefont {Chu}}, \bibinfo {author} {\bibfnamefont {S.}~\bibnamefont {Lederer}},\ and\ \bibinfo {author} {\bibfnamefont {M.}~\bibnamefont {Yi}},\ }\bibfield  {title} {\bibinfo {title} {Nematicity and nematic fluctuations in iron-based superconductors},\ }\href {https://doi.org/10.1038/s41567-022-01833-3} {\bibfield  {journal} {\bibinfo  {journal} {Nature Physics}\ }\textbf {\bibinfo {volume} {18}},\ \bibinfo {pages} {1412} (\bibinfo {year} {2022})}\BibitemShut {NoStop}%
\bibitem [{\citenamefont {Böhmer}\ and\ \citenamefont {Kreisel}(2017)}]{Bohmer_JPCM2018}%
  \BibitemOpen
  \bibfield  {author} {\bibinfo {author} {\bibfnamefont {A.~E.}\ \bibnamefont {Böhmer}}\ and\ \bibinfo {author} {\bibfnamefont {A.}~\bibnamefont {Kreisel}},\ }\bibfield  {title} {\bibinfo {title} {Nematicity, magnetism and superconductivity in {F}e{S}e},\ }\href {https://doi.org/10.1088/1361-648X/aa9caa} {\bibfield  {journal} {\bibinfo  {journal} {Journal of Physics: Condensed Matter}\ }\textbf {\bibinfo {volume} {30}},\ \bibinfo {pages} {023001} (\bibinfo {year} {2017})}\BibitemShut {NoStop}%
\bibitem [{\citenamefont {Kreisel}\ \emph {et~al.}(2020)\citenamefont {Kreisel}, \citenamefont {Hirschfeld},\ and\ \citenamefont {Andersen}}]{Kreisel_Symmetry2020}%
  \BibitemOpen
  \bibfield  {author} {\bibinfo {author} {\bibfnamefont {A.}~\bibnamefont {Kreisel}}, \bibinfo {author} {\bibfnamefont {P.~J.}\ \bibnamefont {Hirschfeld}},\ and\ \bibinfo {author} {\bibfnamefont {B.~M.}\ \bibnamefont {Andersen}},\ }\bibfield  {title} {\bibinfo {title} {On the remarkable superconductivity of {F}e{S}e and its close cousins},\ }\bibfield  {journal} {\bibinfo  {journal} {Symmetry}\ }\textbf {\bibinfo {volume} {12}},\ \href {https://doi.org/10.3390/sym12091402} {10.3390/sym12091402} (\bibinfo {year} {2020})\BibitemShut {NoStop}%
\bibitem [{\citenamefont {Baek}\ \emph {et~al.}(2014)\citenamefont {Baek}, \citenamefont {Efremov}, \citenamefont {Ok}, \citenamefont {Kim}, \citenamefont {van~den Brink},\ and\ \citenamefont {B\"uchner}}]{Baek_NatMat2014}%
  \BibitemOpen
  \bibfield  {author} {\bibinfo {author} {\bibfnamefont {S.-H.}\ \bibnamefont {Baek}}, \bibinfo {author} {\bibfnamefont {D.~V.}\ \bibnamefont {Efremov}}, \bibinfo {author} {\bibfnamefont {J.~M.}\ \bibnamefont {Ok}}, \bibinfo {author} {\bibfnamefont {J.~S.}\ \bibnamefont {Kim}}, \bibinfo {author} {\bibfnamefont {J.}~\bibnamefont {van~den Brink}},\ and\ \bibinfo {author} {\bibfnamefont {B.}~\bibnamefont {B\"uchner}},\ }\bibfield  {title} {\bibinfo {title} {Orbital-driven nematicity in {F}e{S}e},\ }\href {https://doi.org/10.1038/nmat4138} {\bibfield  {journal} {\bibinfo  {journal} {Nature Materials}\ }\textbf {\bibinfo {volume} {14}},\ \bibinfo {pages} {210} (\bibinfo {year} {2014})}\BibitemShut {NoStop}%
\bibitem [{\citenamefont {Su}\ \emph {et~al.}(2015)\citenamefont {Su}, \citenamefont {Liao},\ and\ \citenamefont {Li}}]{Su_JPCM2015}%
  \BibitemOpen
  \bibfield  {author} {\bibinfo {author} {\bibfnamefont {Y.}~\bibnamefont {Su}}, \bibinfo {author} {\bibfnamefont {H.}~\bibnamefont {Liao}},\ and\ \bibinfo {author} {\bibfnamefont {T.}~\bibnamefont {Li}},\ }\bibfield  {title} {\bibinfo {title} {Interplay between magnetism and nematicity in {F}e{S}e},\ }\href {https://doi.org/10.1088/0953-8984/27/10/105702} {\bibfield  {journal} {\bibinfo  {journal} {Journal of Physics: Condensed Matter}\ }\textbf {\bibinfo {volume} {27}},\ \bibinfo {pages} {105702} (\bibinfo {year} {2015})}\BibitemShut {NoStop}%
\bibitem [{\citenamefont {Massat}\ \emph {et~al.}(2016)\citenamefont {Massat}, \citenamefont {Farina}, \citenamefont {Paul}, \citenamefont {Karlsson}, \citenamefont {Strobel}, \citenamefont {Toulemonde}, \citenamefont {M\'easson}, \citenamefont {Cazayous}, \citenamefont {Sacuto}, \citenamefont {Kasahara}, \citenamefont {Shibauchi}, \citenamefont {Matsuda},\ and\ \citenamefont {Gallais}}]{Massat_PNAS2016}%
  \BibitemOpen
  \bibfield  {author} {\bibinfo {author} {\bibfnamefont {P.}~\bibnamefont {Massat}}, \bibinfo {author} {\bibfnamefont {D.}~\bibnamefont {Farina}}, \bibinfo {author} {\bibfnamefont {I.}~\bibnamefont {Paul}}, \bibinfo {author} {\bibfnamefont {S.}~\bibnamefont {Karlsson}}, \bibinfo {author} {\bibfnamefont {P.}~\bibnamefont {Strobel}}, \bibinfo {author} {\bibfnamefont {P.}~\bibnamefont {Toulemonde}}, \bibinfo {author} {\bibfnamefont {M.-A.}\ \bibnamefont {M\'easson}}, \bibinfo {author} {\bibfnamefont {M.}~\bibnamefont {Cazayous}}, \bibinfo {author} {\bibfnamefont {A.}~\bibnamefont {Sacuto}}, \bibinfo {author} {\bibfnamefont {S.}~\bibnamefont {Kasahara}}, \bibinfo {author} {\bibfnamefont {T.}~\bibnamefont {Shibauchi}}, \bibinfo {author} {\bibfnamefont {Y.}~\bibnamefont {Matsuda}},\ and\ \bibinfo {author} {\bibfnamefont {Y.}~\bibnamefont {Gallais}},\ }\bibfield  {title} {\bibinfo {title} {Charge-induced nematicity in {F}e{S}e},\ }\href {https://doi.org/10.1073/pnas.1605769113} {\bibfield  {journal} {\bibinfo
  {journal} {Proceedings of the National Academy of Sciences of the United States of America}\ }\textbf {\bibinfo {volume} {113}},\ \bibinfo {pages} {9177} (\bibinfo {year} {2016})}\BibitemShut {NoStop}%
\bibitem [{\citenamefont {Rahn}\ \emph {et~al.}(2015)\citenamefont {Rahn}, \citenamefont {Ewings}, \citenamefont {Sedlmaier}, \citenamefont {Clarke},\ and\ \citenamefont {Boothroyd}}]{Rahn_PRB2015}%
  \BibitemOpen
  \bibfield  {author} {\bibinfo {author} {\bibfnamefont {M.~C.}\ \bibnamefont {Rahn}}, \bibinfo {author} {\bibfnamefont {R.~A.}\ \bibnamefont {Ewings}}, \bibinfo {author} {\bibfnamefont {S.~J.}\ \bibnamefont {Sedlmaier}}, \bibinfo {author} {\bibfnamefont {S.~J.}\ \bibnamefont {Clarke}},\ and\ \bibinfo {author} {\bibfnamefont {A.~T.}\ \bibnamefont {Boothroyd}},\ }\bibfield  {title} {\bibinfo {title} {Strong $(\ensuremath{\pi},0)$ spin fluctuations in $\ensuremath{\beta}\ensuremath{-}\mathrm{FeSe}$ observed by neutron spectroscopy},\ }\href {https://doi.org/10.1103/PhysRevB.91.180501} {\bibfield  {journal} {\bibinfo  {journal} {Phys. Rev. B}\ }\textbf {\bibinfo {volume} {91}},\ \bibinfo {pages} {180501} (\bibinfo {year} {2015})}\BibitemShut {NoStop}%
\bibitem [{\citenamefont {Wang}\ \emph {et~al.}(2016)\citenamefont {Wang}, \citenamefont {Shen}, \citenamefont {Pan}, \citenamefont {Hao}, \citenamefont {Ma}, \citenamefont {Zhou}, \citenamefont {Steffens}, \citenamefont {Schmalzl}, \citenamefont {Forrest}, \citenamefont {Abdel-Hafiez}, \citenamefont {Chen}, \citenamefont {Chareev}, \citenamefont {Vasiliev}, \citenamefont {Bourges}, \citenamefont {Sidis}, \citenamefont {Cao},\ and\ \citenamefont {Zhao}}]{Wang_NatMat2016}%
  \BibitemOpen
  \bibfield  {author} {\bibinfo {author} {\bibfnamefont {Q.}~\bibnamefont {Wang}}, \bibinfo {author} {\bibfnamefont {Y.}~\bibnamefont {Shen}}, \bibinfo {author} {\bibfnamefont {B.}~\bibnamefont {Pan}}, \bibinfo {author} {\bibfnamefont {Y.}~\bibnamefont {Hao}}, \bibinfo {author} {\bibfnamefont {M.}~\bibnamefont {Ma}}, \bibinfo {author} {\bibfnamefont {F.}~\bibnamefont {Zhou}}, \bibinfo {author} {\bibfnamefont {P.}~\bibnamefont {Steffens}}, \bibinfo {author} {\bibfnamefont {K.}~\bibnamefont {Schmalzl}}, \bibinfo {author} {\bibfnamefont {T.~R.}\ \bibnamefont {Forrest}}, \bibinfo {author} {\bibfnamefont {M.}~\bibnamefont {Abdel-Hafiez}}, \bibinfo {author} {\bibfnamefont {X.}~\bibnamefont {Chen}}, \bibinfo {author} {\bibfnamefont {D.~A.}\ \bibnamefont {Chareev}}, \bibinfo {author} {\bibfnamefont {A.~N.}\ \bibnamefont {Vasiliev}}, \bibinfo {author} {\bibfnamefont {P.}~\bibnamefont {Bourges}}, \bibinfo {author} {\bibfnamefont {Y.}~\bibnamefont {Sidis}}, \bibinfo {author} {\bibfnamefont {H.}~\bibnamefont {Cao}},\
  and\ \bibinfo {author} {\bibfnamefont {J.}~\bibnamefont {Zhao}},\ }\bibfield  {title} {\bibinfo {title} {Strong interplay between stripe spin fluctuations, nematicity and superconductivity in {F}e{S}e},\ }\href {https://doi.org/10.1038/nmat4492a} {\bibfield  {journal} {\bibinfo  {journal} {Nature Materials}\ }\textbf {\bibinfo {volume} {15}},\ \bibinfo {pages} {159} (\bibinfo {year} {2016})}\BibitemShut {NoStop}%
\bibitem [{\citenamefont {Chen}\ \emph {et~al.}(2019)\citenamefont {Chen}, \citenamefont {Chen}, \citenamefont {Kreisel}, \citenamefont {Lu}, \citenamefont {Schneidewind}, \citenamefont {Qiu}, \citenamefont {Park}, \citenamefont {Perring}, \citenamefont {Stewart}, \citenamefont {Cao}, \citenamefont {Zhang}, \citenamefont {Li}, \citenamefont {Rong}, \citenamefont {Wei}, \citenamefont {Andersen}, \citenamefont {Hirschfeld}, \citenamefont {Broholm},\ and\ \citenamefont {Dai}}]{Chen_NatMat2019}%
  \BibitemOpen
  \bibfield  {author} {\bibinfo {author} {\bibfnamefont {T.}~\bibnamefont {Chen}}, \bibinfo {author} {\bibfnamefont {Y.}~\bibnamefont {Chen}}, \bibinfo {author} {\bibfnamefont {A.}~\bibnamefont {Kreisel}}, \bibinfo {author} {\bibfnamefont {X.}~\bibnamefont {Lu}}, \bibinfo {author} {\bibfnamefont {A.}~\bibnamefont {Schneidewind}}, \bibinfo {author} {\bibfnamefont {Y.}~\bibnamefont {Qiu}}, \bibinfo {author} {\bibfnamefont {J.~T.}\ \bibnamefont {Park}}, \bibinfo {author} {\bibfnamefont {T.~G.}\ \bibnamefont {Perring}}, \bibinfo {author} {\bibfnamefont {J.~R.}\ \bibnamefont {Stewart}}, \bibinfo {author} {\bibfnamefont {H.}~\bibnamefont {Cao}}, \bibinfo {author} {\bibfnamefont {R.}~\bibnamefont {Zhang}}, \bibinfo {author} {\bibfnamefont {Y.}~\bibnamefont {Li}}, \bibinfo {author} {\bibfnamefont {Y.}~\bibnamefont {Rong}}, \bibinfo {author} {\bibfnamefont {Y.}~\bibnamefont {Wei}}, \bibinfo {author} {\bibfnamefont {B.~M.}\ \bibnamefont {Andersen}}, \bibinfo {author} {\bibfnamefont {P.~J.}\ \bibnamefont {Hirschfeld}},
  \bibinfo {author} {\bibfnamefont {C.}~\bibnamefont {Broholm}},\ and\ \bibinfo {author} {\bibfnamefont {P.}~\bibnamefont {Dai}},\ }\bibfield  {title} {\bibinfo {title} {Anisotropic spin fluctuations in detwinned {F}e{S}e},\ }\href {https://doi.org/0.1038/s41563-019-0369-5} {\bibfield  {journal} {\bibinfo  {journal} {Nature Materials}\ }\textbf {\bibinfo {volume} {18}},\ \bibinfo {pages} {706} (\bibinfo {year} {2019})}\BibitemShut {NoStop}%
\bibitem [{\citenamefont {Zhou}\ \emph {et~al.}(2020)\citenamefont {Zhou}, \citenamefont {Scherer}, \citenamefont {Mayaffre}, \citenamefont {Toulemonde}, \citenamefont {Ma}, \citenamefont {Li}, \citenamefont {Andersen},\ and\ \citenamefont {Julien}}]{Zhou_npjQM2020}%
  \BibitemOpen
  \bibfield  {author} {\bibinfo {author} {\bibfnamefont {R.}~\bibnamefont {Zhou}}, \bibinfo {author} {\bibfnamefont {D.~D.}\ \bibnamefont {Scherer}}, \bibinfo {author} {\bibfnamefont {H.}~\bibnamefont {Mayaffre}}, \bibinfo {author} {\bibfnamefont {P.}~\bibnamefont {Toulemonde}}, \bibinfo {author} {\bibfnamefont {M.}~\bibnamefont {Ma}}, \bibinfo {author} {\bibfnamefont {Y.}~\bibnamefont {Li}}, \bibinfo {author} {\bibfnamefont {B.~M.}\ \bibnamefont {Andersen}},\ and\ \bibinfo {author} {\bibfnamefont {M.-H.}\ \bibnamefont {Julien}},\ }\bibfield  {title} {\bibinfo {title} {Singular magnetic anisotropy in the nematic phase of {F}e{S}e},\ }\href {https://doi.org/10.1038/s41535-020-00295-1} {\bibfield  {journal} {\bibinfo  {journal} {npj Quantum Materials}\ }\textbf {\bibinfo {volume} {5}},\ \bibinfo {pages} {93} (\bibinfo {year} {2020})}\BibitemShut {NoStop}%
\bibitem [{\citenamefont {Fanfarillo}\ \emph {et~al.}(2015)\citenamefont {Fanfarillo}, \citenamefont {Cortijo},\ and\ \citenamefont {Valenzuela}}]{Fanfarillo_PRB2015}%
  \BibitemOpen
  \bibfield  {author} {\bibinfo {author} {\bibfnamefont {L.}~\bibnamefont {Fanfarillo}}, \bibinfo {author} {\bibfnamefont {A.}~\bibnamefont {Cortijo}},\ and\ \bibinfo {author} {\bibfnamefont {B.}~\bibnamefont {Valenzuela}},\ }\bibfield  {title} {\bibinfo {title} {Spin-orbital interplay and topology in the nematic phase of iron pnictides},\ }\href {https://doi.org/10.1103/PhysRevB.91.214515} {\bibfield  {journal} {\bibinfo  {journal} {Phys. Rev. B}\ }\textbf {\bibinfo {volume} {91}},\ \bibinfo {pages} {214515} (\bibinfo {year} {2015})}\BibitemShut {NoStop}%
\bibitem [{\citenamefont {Glasbrenner}\ \emph {et~al.}(2015)\citenamefont {Glasbrenner}, \citenamefont {Mazin}, \citenamefont {Jeschke}, \citenamefont {Hirschfeld}, \citenamefont {Fernandes},\ and\ \citenamefont {Valent\'i}}]{Glasbrenner_NatPhys2015}%
  \BibitemOpen
  \bibfield  {author} {\bibinfo {author} {\bibfnamefont {J.~K.}\ \bibnamefont {Glasbrenner}}, \bibinfo {author} {\bibfnamefont {I.~I.}\ \bibnamefont {Mazin}}, \bibinfo {author} {\bibfnamefont {H.~O.}\ \bibnamefont {Jeschke}}, \bibinfo {author} {\bibfnamefont {P.~J.}\ \bibnamefont {Hirschfeld}}, \bibinfo {author} {\bibfnamefont {R.~M.}\ \bibnamefont {Fernandes}},\ and\ \bibinfo {author} {\bibfnamefont {R.}~\bibnamefont {Valent\'i}},\ }\bibfield  {title} {\bibinfo {title} {Effect of magnetic frustration on nematicity and superconductivity in iron chalcogenides},\ }\href {https://doi.org/10.1038/nphys3434} {\bibfield  {journal} {\bibinfo  {journal} {Nature Physics}\ }\textbf {\bibinfo {volume} {11}},\ \bibinfo {pages} {953} (\bibinfo {year} {2015})}\BibitemShut {NoStop}%
\bibitem [{\citenamefont {Christensen}\ \emph {et~al.}(2016)\citenamefont {Christensen}, \citenamefont {Kang}, \citenamefont {Andersen},\ and\ \citenamefont {Fernandes}}]{Christensen_PRB2016}%
  \BibitemOpen
  \bibfield  {author} {\bibinfo {author} {\bibfnamefont {M.~H.}\ \bibnamefont {Christensen}}, \bibinfo {author} {\bibfnamefont {J.-H.}\ \bibnamefont {Kang}}, \bibinfo {author} {\bibfnamefont {B.~M.}\ \bibnamefont {Andersen}},\ and\ \bibinfo {author} {\bibfnamefont {R.~M.}\ \bibnamefont {Fernandes}},\ }\bibfield  {title} {\bibinfo {title} {Spin-driven nematic instability of the multi-orbital hubbard model: Application to iron-based superconductors},\ }\href {https://doi.org/10.1103/PhysRevB.93.085136} {\bibfield  {journal} {\bibinfo  {journal} {Physical Review B}\ }\textbf {\bibinfo {volume} {93}},\ \bibinfo {pages} {085136} (\bibinfo {year} {2016})}\BibitemShut {NoStop}%
\bibitem [{\citenamefont {Fernandes}\ and\ \citenamefont {Chubukov}(2017)}]{Fernandes_Rep2017}%
  \BibitemOpen
  \bibfield  {author} {\bibinfo {author} {\bibfnamefont {R.~M.}\ \bibnamefont {Fernandes}}\ and\ \bibinfo {author} {\bibfnamefont {A.~V.}\ \bibnamefont {Chubukov}},\ }\bibfield  {title} {\bibinfo {title} {Low-energy microscopic models for iron-based superconductors: a review},\ }\href {https://doi.org/10.1088/1361-6633/80/1/014503} {\bibfield  {journal} {\bibinfo  {journal} {Reports on Progress in Physics}\ }\textbf {\bibinfo {volume} {80}},\ \bibinfo {pages} {014503} (\bibinfo {year} {2017})}\BibitemShut {NoStop}%
\bibitem [{\citenamefont {Kreisel}\ \emph {et~al.}(2022)\citenamefont {Kreisel}, \citenamefont {Hirschfeld},\ and\ \citenamefont {Andersen}}]{Kreisel_Frontier2022}%
  \BibitemOpen
  \bibfield  {author} {\bibinfo {author} {\bibfnamefont {A.}~\bibnamefont {Kreisel}}, \bibinfo {author} {\bibfnamefont {P.~J.}\ \bibnamefont {Hirschfeld}},\ and\ \bibinfo {author} {\bibfnamefont {B.~M.}\ \bibnamefont {Andersen}},\ }\bibfield  {title} {\bibinfo {title} {Theory of spin-excitation anisotropy in the nematic phase of {F}e{S}e obtained from rixs measurements},\ }\bibfield  {journal} {\bibinfo  {journal} {Frontiers in Physics}\ }\textbf {\bibinfo {volume} {10}},\ \href {https://doi.org/10.3389/fphy.2022.859424} {10.3389/fphy.2022.859424} (\bibinfo {year} {2022})\BibitemShut {NoStop}%
\bibitem [{\citenamefont {Fanfarillo}\ \emph {et~al.}(2016)\citenamefont {Fanfarillo}, \citenamefont {Mansart}, \citenamefont {Toulemonde}, \citenamefont {Cercellier}, \citenamefont {F\`evre}, \citenamefont {Bertran}, \citenamefont {Valenzuela}, \citenamefont {Benfatto},\ and\ \citenamefont {Brouet}}]{Fanfarillo_PRB2016}%
  \BibitemOpen
  \bibfield  {author} {\bibinfo {author} {\bibfnamefont {L.}~\bibnamefont {Fanfarillo}}, \bibinfo {author} {\bibfnamefont {J.}~\bibnamefont {Mansart}}, \bibinfo {author} {\bibfnamefont {P.}~\bibnamefont {Toulemonde}}, \bibinfo {author} {\bibfnamefont {H.}~\bibnamefont {Cercellier}}, \bibinfo {author} {\bibfnamefont {P.~L.}\ \bibnamefont {F\`evre}}, \bibinfo {author} {\bibfnamefont {F.}~\bibnamefont {Bertran}}, \bibinfo {author} {\bibfnamefont {B.}~\bibnamefont {Valenzuela}}, \bibinfo {author} {\bibfnamefont {L.}~\bibnamefont {Benfatto}},\ and\ \bibinfo {author} {\bibfnamefont {V.}~\bibnamefont {Brouet}},\ }\bibfield  {title} {\bibinfo {title} {Orbital-dependent fermi surface shrinking as a fingerprint of nematicity in {F}e{S}e},\ }\href {https://doi.org/10.1103/PhysRevB.94.155138} {\bibfield  {journal} {\bibinfo  {journal} {Physical Review B}\ }\textbf {\bibinfo {volume} {94}},\ \bibinfo {pages} {155138} (\bibinfo {year} {2016})}\BibitemShut {NoStop}%
\bibitem [{\citenamefont {Fanfarillo}\ \emph {et~al.}(2018)\citenamefont {Fanfarillo}, \citenamefont {Benfatto},\ and\ \citenamefont {Valenzuela}}]{Fanfarillo_PRB2018}%
  \BibitemOpen
  \bibfield  {author} {\bibinfo {author} {\bibfnamefont {L.}~\bibnamefont {Fanfarillo}}, \bibinfo {author} {\bibfnamefont {L.}~\bibnamefont {Benfatto}},\ and\ \bibinfo {author} {\bibfnamefont {B.}~\bibnamefont {Valenzuela}},\ }\bibfield  {title} {\bibinfo {title} {Orbital mismatch boosting nematic instability in iron-based superconductors},\ }\href {https://doi.org/10.1103/PhysRevB.97.121109} {\bibfield  {journal} {\bibinfo  {journal} {Phys. Rev. B}\ }\textbf {\bibinfo {volume} {97}},\ \bibinfo {pages} {121109} (\bibinfo {year} {2018})}\BibitemShut {NoStop}%
\bibitem [{\citenamefont {Chu}\ \emph {et~al.}(2012)\citenamefont {Chu}, \citenamefont {Kuo}, \citenamefont {Analytis},\ and\ \citenamefont {Fisher}}]{ChuScience}%
  \BibitemOpen
  \bibfield  {author} {\bibinfo {author} {\bibfnamefont {J.-H.}\ \bibnamefont {Chu}}, \bibinfo {author} {\bibfnamefont {H.-H.}\ \bibnamefont {Kuo}}, \bibinfo {author} {\bibfnamefont {J.~G.}\ \bibnamefont {Analytis}},\ and\ \bibinfo {author} {\bibfnamefont {I.~R.}\ \bibnamefont {Fisher}},\ }\bibfield  {title} {\bibinfo {title} {Divergent nematic susceptibility in an iron arsenide superconductor},\ }\href {https://doi.org/10.1126/science.1221713} {\bibfield  {journal} {\bibinfo  {journal} {Science}\ }\textbf {\bibinfo {volume} {337}},\ \bibinfo {pages} {710} (\bibinfo {year} {2012})},\ \Eprint {https://arxiv.org/abs/https://www.science.org/doi/pdf/10.1126/science.1221713} {https://www.science.org/doi/pdf/10.1126/science.1221713} \BibitemShut {NoStop}%
\bibitem [{\citenamefont {Kuo}\ \emph {et~al.}(2016)\citenamefont {Kuo}, \citenamefont {Chu}, \citenamefont {Palmstrom}, \citenamefont {Kivelson},\ and\ \citenamefont {Fisher}}]{KuoScience}%
  \BibitemOpen
  \bibfield  {author} {\bibinfo {author} {\bibfnamefont {H.-H.}\ \bibnamefont {Kuo}}, \bibinfo {author} {\bibfnamefont {J.-H.}\ \bibnamefont {Chu}}, \bibinfo {author} {\bibfnamefont {J.~C.}\ \bibnamefont {Palmstrom}}, \bibinfo {author} {\bibfnamefont {S.~A.}\ \bibnamefont {Kivelson}},\ and\ \bibinfo {author} {\bibfnamefont {I.~R.}\ \bibnamefont {Fisher}},\ }\bibfield  {title} {\bibinfo {title} {Ubiquitous signatures of nematic quantum criticality in optimally doped {F}e-based superconductors},\ }\href {https://doi.org/10.1126/science.aab0103} {\bibfield  {journal} {\bibinfo  {journal} {Science}\ }\textbf {\bibinfo {volume} {352}},\ \bibinfo {pages} {958} (\bibinfo {year} {2016})},\ \Eprint {https://arxiv.org/abs/https://www.science.org/doi/pdf/10.1126/science.aab0103} {https://www.science.org/doi/pdf/10.1126/science.aab0103} \BibitemShut {NoStop}%
\bibitem [{\citenamefont {Wissmann}\ \emph {et~al.}(2022)\citenamefont {Wissmann}, \citenamefont {Caglieris}, \citenamefont {Hong}, \citenamefont {Aswartham}, \citenamefont {Vorobyova}, \citenamefont {Morozov}, \citenamefont {B\"uchner},\ and\ \citenamefont {Hess}}]{PhysRevB.106.054508}%
  \BibitemOpen
  \bibfield  {author} {\bibinfo {author} {\bibfnamefont {M.}~\bibnamefont {Wissmann}}, \bibinfo {author} {\bibfnamefont {F.}~\bibnamefont {Caglieris}}, \bibinfo {author} {\bibfnamefont {X.}~\bibnamefont {Hong}}, \bibinfo {author} {\bibfnamefont {S.}~\bibnamefont {Aswartham}}, \bibinfo {author} {\bibfnamefont {A.}~\bibnamefont {Vorobyova}}, \bibinfo {author} {\bibfnamefont {I.}~\bibnamefont {Morozov}}, \bibinfo {author} {\bibfnamefont {B.}~\bibnamefont {B\"uchner}},\ and\ \bibinfo {author} {\bibfnamefont {C.}~\bibnamefont {Hess}},\ }\bibfield  {title} {\bibinfo {title} {Absence of nematic instability in {L}i{F}e{A}s},\ }\href {https://doi.org/10.1103/PhysRevB.106.054508} {\bibfield  {journal} {\bibinfo  {journal} {Phys. Rev. B}\ }\textbf {\bibinfo {volume} {106}},\ \bibinfo {pages} {054508} (\bibinfo {year} {2022})}\BibitemShut {NoStop}%
\bibitem [{\citenamefont {Hong}\ \emph {et~al.}(2022)\citenamefont {Hong}, \citenamefont {Sykora}, \citenamefont {Caglieris}, \citenamefont {Behnami}, \citenamefont {Morozov}, \citenamefont {Aswartham}, \citenamefont {Grinenko}, \citenamefont {Kihou}, \citenamefont {Lee}, \citenamefont {Büchner},\ and\ \citenamefont {Hess}}]{122XC}%
  \BibitemOpen
  \bibfield  {author} {\bibinfo {author} {\bibfnamefont {X.}~\bibnamefont {Hong}}, \bibinfo {author} {\bibfnamefont {S.}~\bibnamefont {Sykora}}, \bibinfo {author} {\bibfnamefont {F.}~\bibnamefont {Caglieris}}, \bibinfo {author} {\bibfnamefont {M.}~\bibnamefont {Behnami}}, \bibinfo {author} {\bibfnamefont {I.}~\bibnamefont {Morozov}}, \bibinfo {author} {\bibfnamefont {S.}~\bibnamefont {Aswartham}}, \bibinfo {author} {\bibfnamefont {V.}~\bibnamefont {Grinenko}}, \bibinfo {author} {\bibfnamefont {K.}~\bibnamefont {Kihou}}, \bibinfo {author} {\bibfnamefont {C.-H.}\ \bibnamefont {Lee}}, \bibinfo {author} {\bibfnamefont {B.}~\bibnamefont {Büchner}},\ and\ \bibinfo {author} {\bibfnamefont {C.}~\bibnamefont {Hess}},\ }\bibfield  {title} {\bibinfo {title} {Elastoresistivity of heavily hole-doped 122 iron pnictide superconductors},\ }\bibfield  {journal} {\bibinfo  {journal} {Frontiers in Physics}\ }\textbf {\bibinfo {volume} {10}},\ \href {https://doi.org/10.3389/fphy.2022.853717} {10.3389/fphy.2022.853717} (\bibinfo
  {year} {2022})\BibitemShut {NoStop}%
\bibitem [{\citenamefont {Tanatar}\ \emph {et~al.}(2016)\citenamefont {Tanatar}, \citenamefont {B\"ohmer}, \citenamefont {Timmons}, \citenamefont {Sch\"utt}, \citenamefont {Drachuck}, \citenamefont {Taufour}, \citenamefont {Kothapalli}, \citenamefont {Kreyssig}, \citenamefont {Bud'ko}, \citenamefont {Canfield}, \citenamefont {Fernandes},\ and\ \citenamefont {Prozorov}}]{FeSePhysRevLett.117.127001}%
  \BibitemOpen
  \bibfield  {author} {\bibinfo {author} {\bibfnamefont {M.~A.}\ \bibnamefont {Tanatar}}, \bibinfo {author} {\bibfnamefont {A.~E.}\ \bibnamefont {B\"ohmer}}, \bibinfo {author} {\bibfnamefont {E.~I.}\ \bibnamefont {Timmons}}, \bibinfo {author} {\bibfnamefont {M.}~\bibnamefont {Sch\"utt}}, \bibinfo {author} {\bibfnamefont {G.}~\bibnamefont {Drachuck}}, \bibinfo {author} {\bibfnamefont {V.}~\bibnamefont {Taufour}}, \bibinfo {author} {\bibfnamefont {K.}~\bibnamefont {Kothapalli}}, \bibinfo {author} {\bibfnamefont {A.}~\bibnamefont {Kreyssig}}, \bibinfo {author} {\bibfnamefont {S.~L.}\ \bibnamefont {Bud'ko}}, \bibinfo {author} {\bibfnamefont {P.~C.}\ \bibnamefont {Canfield}}, \bibinfo {author} {\bibfnamefont {R.~M.}\ \bibnamefont {Fernandes}},\ and\ \bibinfo {author} {\bibfnamefont {R.}~\bibnamefont {Prozorov}},\ }\bibfield  {title} {\bibinfo {title} {Origin of the resistivity anisotropy in the nematic phase of {F}e{S}e},\ }\href {https://doi.org/10.1103/PhysRevLett.117.127001} {\bibfield  {journal} {\bibinfo
  {journal} {Phys. Rev. Lett.}\ }\textbf {\bibinfo {volume} {117}},\ \bibinfo {pages} {127001} (\bibinfo {year} {2016})}\BibitemShut {NoStop}%
\bibitem [{\citenamefont {Bartlett}\ \emph {et~al.}(2021)\citenamefont {Bartlett}, \citenamefont {Steppke}, \citenamefont {Hosoi}, \citenamefont {Noad}, \citenamefont {Park}, \citenamefont {Timm}, \citenamefont {Shibauchi}, \citenamefont {Mackenzie},\ and\ \citenamefont {Hicks}}]{FeSePhysRevX.11.021038}%
  \BibitemOpen
  \bibfield  {author} {\bibinfo {author} {\bibfnamefont {J.~M.}\ \bibnamefont {Bartlett}}, \bibinfo {author} {\bibfnamefont {A.}~\bibnamefont {Steppke}}, \bibinfo {author} {\bibfnamefont {S.}~\bibnamefont {Hosoi}}, \bibinfo {author} {\bibfnamefont {H.}~\bibnamefont {Noad}}, \bibinfo {author} {\bibfnamefont {J.}~\bibnamefont {Park}}, \bibinfo {author} {\bibfnamefont {C.}~\bibnamefont {Timm}}, \bibinfo {author} {\bibfnamefont {T.}~\bibnamefont {Shibauchi}}, \bibinfo {author} {\bibfnamefont {A.~P.}\ \bibnamefont {Mackenzie}},\ and\ \bibinfo {author} {\bibfnamefont {C.~W.}\ \bibnamefont {Hicks}},\ }\bibfield  {title} {\bibinfo {title} {Relationship between transport anisotropy and nematicity in {F}e{S}e},\ }\href {https://doi.org/10.1103/PhysRevX.11.021038} {\bibfield  {journal} {\bibinfo  {journal} {Phys. Rev. X}\ }\textbf {\bibinfo {volume} {11}},\ \bibinfo {pages} {021038} (\bibinfo {year} {2021})}\BibitemShut {NoStop}%
\bibitem [{\citenamefont {Hong}\ \emph {et~al.}(2020)\citenamefont {Hong}, \citenamefont {Caglieris}, \citenamefont {Kappenberger}, \citenamefont {Wurmehl}, \citenamefont {Aswartham}, \citenamefont {Scaravaggi}, \citenamefont {Lepucki}, \citenamefont {Wolter}, \citenamefont {Grafe}, \citenamefont {B\"uchner},\ and\ \citenamefont {Hess}}]{PhysRevLett.125.067001}%
  \BibitemOpen
  \bibfield  {author} {\bibinfo {author} {\bibfnamefont {X.}~\bibnamefont {Hong}}, \bibinfo {author} {\bibfnamefont {F.}~\bibnamefont {Caglieris}}, \bibinfo {author} {\bibfnamefont {R.}~\bibnamefont {Kappenberger}}, \bibinfo {author} {\bibfnamefont {S.}~\bibnamefont {Wurmehl}}, \bibinfo {author} {\bibfnamefont {S.}~\bibnamefont {Aswartham}}, \bibinfo {author} {\bibfnamefont {F.}~\bibnamefont {Scaravaggi}}, \bibinfo {author} {\bibfnamefont {P.}~\bibnamefont {Lepucki}}, \bibinfo {author} {\bibfnamefont {A.~U.~B.}\ \bibnamefont {Wolter}}, \bibinfo {author} {\bibfnamefont {H.-J.}\ \bibnamefont {Grafe}}, \bibinfo {author} {\bibfnamefont {B.}~\bibnamefont {B\"uchner}},\ and\ \bibinfo {author} {\bibfnamefont {C.}~\bibnamefont {Hess}},\ }\bibfield  {title} {\bibinfo {title} {Evolution of the nematic susceptibility in ${\mathrm{lafe}}_{1\ensuremath{-}x}{\mathrm{co}}_{x}\mathrm{AsO}$},\ }\href {https://doi.org/10.1103/PhysRevLett.125.067001} {\bibfield  {journal} {\bibinfo  {journal} {Phys. Rev. Lett.}\ }\textbf
  {\bibinfo {volume} {125}},\ \bibinfo {pages} {067001} (\bibinfo {year} {2020})}\BibitemShut {NoStop}%
\bibitem [{\citenamefont {Caglieris}\ \emph {et~al.}(2021)\citenamefont {Caglieris}, \citenamefont {Wuttke}, \citenamefont {Hong}, \citenamefont {Sykora}, \citenamefont {Kappenberger}, \citenamefont {Aswartham}, \citenamefont {Wurmehl}, \citenamefont {Büchner},\ and\ \citenamefont {Hess}}]{NpjCcaglieris}%
  \BibitemOpen
  \bibfield  {author} {\bibinfo {author} {\bibfnamefont {F.}~\bibnamefont {Caglieris}}, \bibinfo {author} {\bibfnamefont {C.}~\bibnamefont {Wuttke}}, \bibinfo {author} {\bibfnamefont {X.}~\bibnamefont {Hong}}, \bibinfo {author} {\bibfnamefont {S.}~\bibnamefont {Sykora}}, \bibinfo {author} {\bibfnamefont {R.}~\bibnamefont {Kappenberger}}, \bibinfo {author} {\bibfnamefont {S.}~\bibnamefont {Aswartham}}, \bibinfo {author} {\bibfnamefont {S.}~\bibnamefont {Wurmehl}}, \bibinfo {author} {\bibfnamefont {B.}~\bibnamefont {Büchner}},\ and\ \bibinfo {author} {\bibfnamefont {C.}~\bibnamefont {Hess}},\ }\bibfield  {title} {\bibinfo {title} {Strain derivative of thermoelectric properties as a sensitive probe for nematicity},\ }\href {https://doi.org/10.1038/s41535-021-00324-7} {\bibfield  {journal} {\bibinfo  {journal} {npj Quantum Materials}\ }\textbf {\bibinfo {volume} {6}},\ \bibinfo {pages} {27} (\bibinfo {year} {2021})}\BibitemShut {NoStop}%
\bibitem [{\citenamefont {Baek}\ \emph {et~al.}(2020)\citenamefont {Baek}, \citenamefont {Ok}, \citenamefont {Kim}, \citenamefont {Aswartham}, \citenamefont {Morozov}, \citenamefont {Chareev}, \citenamefont {Urata}, \citenamefont {Tanigaki}, \citenamefont {B\"uchner},\ and\ \citenamefont {Efremov}}]{Baek_2020}%
  \BibitemOpen
  \bibfield  {author} {\bibinfo {author} {\bibfnamefont {S.-H.}\ \bibnamefont {Baek}}, \bibinfo {author} {\bibfnamefont {J.}~\bibnamefont {Ok}}, \bibinfo {author} {\bibfnamefont {J.}~\bibnamefont {Kim}}, \bibinfo {author} {\bibfnamefont {S.}~\bibnamefont {Aswartham}}, \bibinfo {author} {\bibfnamefont {I.}~\bibnamefont {Morozov}}, \bibinfo {author} {\bibfnamefont {D.}~\bibnamefont {Chareev}}, \bibinfo {author} {\bibfnamefont {T.}~\bibnamefont {Urata}}, \bibinfo {author} {\bibfnamefont {Y.}~\bibnamefont {Tanigaki}, \bibfnamefont {K.~Tanabe}}, \bibinfo {author} {\bibfnamefont {B.}~\bibnamefont {B\"uchner}},\ and\ \bibinfo {author} {\bibfnamefont {D.}~\bibnamefont {Efremov}},\ }\bibfield  {title} {\bibinfo {title} {Separate tuning of nematicity and spin fluctuations to unravel the origin of superconductivity in {F}e{S}e},\ }\href {https://doi.org/https://doi.org/10.1038/s41535-020-0211-y} {\bibfield  {journal} {\bibinfo  {journal} {npj Quantum Materials}\ }\textbf {\bibinfo {volume} {5}},\ \bibinfo {pages} {8}
  (\bibinfo {year} {2020})}\BibitemShut {NoStop}%
\bibitem [{\citenamefont {Kasahara}\ \emph {et~al.}(2014)\citenamefont {Kasahara}, \citenamefont {Watashige}, \citenamefont {Hanaguri}, \citenamefont {Kohsaka}, \citenamefont {Yamashita}, \citenamefont {Shimoyama}, \citenamefont {Mizukami}, \citenamefont {Endo}, \citenamefont {Ikeda}, \citenamefont {Aoyama}, \citenamefont {Terashima}, \citenamefont {Uji}, \citenamefont {Wolf}, \citenamefont {von Löhneysen}, \citenamefont {Shibauchi},\ and\ \citenamefont {Matsuda}}]{doi:10.1073/pnas.1413477111}%
  \BibitemOpen
  \bibfield  {author} {\bibinfo {author} {\bibfnamefont {S.}~\bibnamefont {Kasahara}}, \bibinfo {author} {\bibfnamefont {T.}~\bibnamefont {Watashige}}, \bibinfo {author} {\bibfnamefont {T.}~\bibnamefont {Hanaguri}}, \bibinfo {author} {\bibfnamefont {Y.}~\bibnamefont {Kohsaka}}, \bibinfo {author} {\bibfnamefont {T.}~\bibnamefont {Yamashita}}, \bibinfo {author} {\bibfnamefont {Y.}~\bibnamefont {Shimoyama}}, \bibinfo {author} {\bibfnamefont {Y.}~\bibnamefont {Mizukami}}, \bibinfo {author} {\bibfnamefont {R.}~\bibnamefont {Endo}}, \bibinfo {author} {\bibfnamefont {H.}~\bibnamefont {Ikeda}}, \bibinfo {author} {\bibfnamefont {K.}~\bibnamefont {Aoyama}}, \bibinfo {author} {\bibfnamefont {T.}~\bibnamefont {Terashima}}, \bibinfo {author} {\bibfnamefont {S.}~\bibnamefont {Uji}}, \bibinfo {author} {\bibfnamefont {T.}~\bibnamefont {Wolf}}, \bibinfo {author} {\bibfnamefont {H.}~\bibnamefont {von Löhneysen}}, \bibinfo {author} {\bibfnamefont {T.}~\bibnamefont {Shibauchi}},\ and\ \bibinfo {author} {\bibfnamefont
  {Y.}~\bibnamefont {Matsuda}},\ }\bibfield  {title} {\bibinfo {title} {Field-induced superconducting phase of fese in the bcs-bec cross-over},\ }\href {https://doi.org/10.1073/pnas.1413477111} {\bibfield  {journal} {\bibinfo  {journal} {Proceedings of the National Academy of Sciences}\ }\textbf {\bibinfo {volume} {111}},\ \bibinfo {pages} {16309} (\bibinfo {year} {2014})},\ \Eprint {https://arxiv.org/abs/https://www.pnas.org/doi/pdf/10.1073/pnas.1413477111} {https://www.pnas.org/doi/pdf/10.1073/pnas.1413477111} \BibitemShut {NoStop}%
\bibitem [{\citenamefont {\ifmmode~\check{C}\else \v{C}\fi{}ulo}\ \emph {et~al.}(2021)\citenamefont {\ifmmode~\check{C}\else \v{C}\fi{}ulo}, \citenamefont {Berben}, \citenamefont {Hsu}, \citenamefont {Ayres}, \citenamefont {Hinlopen}, \citenamefont {Kasahara}, \citenamefont {Matsuda}, \citenamefont {Shibauchi},\ and\ \citenamefont {Hussey}}]{PhysRevResearch.3.023069}%
  \BibitemOpen
  \bibfield  {author} {\bibinfo {author} {\bibfnamefont {M.}~\bibnamefont {\ifmmode~\check{C}\else \v{C}\fi{}ulo}}, \bibinfo {author} {\bibfnamefont {M.}~\bibnamefont {Berben}}, \bibinfo {author} {\bibfnamefont {Y.-T.}\ \bibnamefont {Hsu}}, \bibinfo {author} {\bibfnamefont {J.}~\bibnamefont {Ayres}}, \bibinfo {author} {\bibfnamefont {R.~D.~H.}\ \bibnamefont {Hinlopen}}, \bibinfo {author} {\bibfnamefont {S.}~\bibnamefont {Kasahara}}, \bibinfo {author} {\bibfnamefont {Y.}~\bibnamefont {Matsuda}}, \bibinfo {author} {\bibfnamefont {T.}~\bibnamefont {Shibauchi}},\ and\ \bibinfo {author} {\bibfnamefont {N.~E.}\ \bibnamefont {Hussey}},\ }\bibfield  {title} {\bibinfo {title} {Putative hall response of the strange metal component in $\mathrm{Fe}{\mathrm{se}}_{1\ensuremath{-}x}{\mathrm{s}}_{x}$},\ }\href {https://doi.org/10.1103/PhysRevResearch.3.023069} {\bibfield  {journal} {\bibinfo  {journal} {Phys. Rev. Res.}\ }\textbf {\bibinfo {volume} {3}},\ \bibinfo {pages} {023069} (\bibinfo {year} {2021})}\BibitemShut
  {NoStop}%
\bibitem [{\citenamefont {Caglieris}\ \emph {et~al.}(2012)\citenamefont {Caglieris}, \citenamefont {Ricci}, \citenamefont {Lamura}, \citenamefont {Martinelli}, \citenamefont {Palenzona}, \citenamefont {Pallecchi}, \citenamefont {Sala}, \citenamefont {Profeta},\ and\ \citenamefont {Putti}}]{Caglieris01102012}%
  \BibitemOpen
  \bibfield  {author} {\bibinfo {author} {\bibfnamefont {F.}~\bibnamefont {Caglieris}}, \bibinfo {author} {\bibfnamefont {F.}~\bibnamefont {Ricci}}, \bibinfo {author} {\bibfnamefont {G.}~\bibnamefont {Lamura}}, \bibinfo {author} {\bibfnamefont {A.}~\bibnamefont {Martinelli}}, \bibinfo {author} {\bibfnamefont {A.}~\bibnamefont {Palenzona}}, \bibinfo {author} {\bibfnamefont {I.}~\bibnamefont {Pallecchi}}, \bibinfo {author} {\bibfnamefont {A.}~\bibnamefont {Sala}}, \bibinfo {author} {\bibfnamefont {G.}~\bibnamefont {Profeta}},\ and\ \bibinfo {author} {\bibfnamefont {M.}~\bibnamefont {Putti}},\ }\bibfield  {title} {\bibinfo {title} {Theoretical and experimental investigation of magnetotransport in iron chalcogenides},\ }\href {https://doi.org/10.1088/1468-6996/13/5/054402} {\bibfield  {journal} {\bibinfo  {journal} {Science and Technology of Advanced Materials}\ }\textbf {\bibinfo {volume} {13}},\ \bibinfo {pages} {054402} (\bibinfo {year} {2012})},\ \bibinfo {note} {pMID: 27877520},\ \Eprint
  {https://arxiv.org/abs/https://doi.org/10.1088/1468-6996/13/5/054402} {https://doi.org/10.1088/1468-6996/13/5/054402} \BibitemShut {NoStop}%
\bibitem [{\citenamefont {Fern\'andez-Mart\'{\i}n}\ \emph {et~al.}(2019)\citenamefont {Fern\'andez-Mart\'{\i}n}, \citenamefont {Fanfarillo}, \citenamefont {Benfatto},\ and\ \citenamefont {Valenzuela}}]{Fernandez_PRB2019}%
  \BibitemOpen
  \bibfield  {author} {\bibinfo {author} {\bibfnamefont {R.}~\bibnamefont {Fern\'andez-Mart\'{\i}n}}, \bibinfo {author} {\bibfnamefont {L.}~\bibnamefont {Fanfarillo}}, \bibinfo {author} {\bibfnamefont {L.}~\bibnamefont {Benfatto}},\ and\ \bibinfo {author} {\bibfnamefont {B.}~\bibnamefont {Valenzuela}},\ }\bibfield  {title} {\bibinfo {title} {Anisotropy of the dc conductivity due to orbital-selective spin fluctuations in the nematic phase of iron superconductors},\ }\href {https://doi.org/10.1103/PhysRevB.99.155117} {\bibfield  {journal} {\bibinfo  {journal} {Phys. Rev. B}\ }\textbf {\bibinfo {volume} {99}},\ \bibinfo {pages} {155117} (\bibinfo {year} {2019})}\BibitemShut {NoStop}%
\bibitem [{\citenamefont {Marciani}\ and\ \citenamefont {Benfatto}(2022)}]{Marciani_PRB2022}%
  \BibitemOpen
  \bibfield  {author} {\bibinfo {author} {\bibfnamefont {M.}~\bibnamefont {Marciani}}\ and\ \bibinfo {author} {\bibfnamefont {L.}~\bibnamefont {Benfatto}},\ }\bibfield  {title} {\bibinfo {title} {Resistivity anisotropy from the multiorbital boltzmann equation in nematic {F}e{S}e},\ }\href {https://doi.org/10.1103/PhysRevB.106.045102} {\bibfield  {journal} {\bibinfo  {journal} {Phys. Rev. B}\ }\textbf {\bibinfo {volume} {106}},\ \bibinfo {pages} {045102} (\bibinfo {year} {2022})}\BibitemShut {NoStop}%
\bibitem [{\citenamefont {Jones}\ and\ \citenamefont {Zener}(1934)}]{jonesZener34}%
  \BibitemOpen
  \bibfield  {author} {\bibinfo {author} {\bibfnamefont {H.}~\bibnamefont {Jones}}\ and\ \bibinfo {author} {\bibfnamefont {C.}~\bibnamefont {Zener}},\ }\bibfield  {title} {\bibinfo {title} {Iron pnictides and chalcogenides: a new paradigm for superconductivity},\ }\href {https://doi.org/https://doi.org/10.1098/rspa.1934.0095} {\bibfield  {journal} {\bibinfo  {journal} {Proc. Royal. Soc.}\ }\textbf {\bibinfo {volume} {A145}},\ \bibinfo {pages} {268} (\bibinfo {year} {1934})}\BibitemShut {NoStop}%
\bibitem [{\citenamefont {Chambers}(1952)}]{Chambers_ProcPhysSocA1952}%
  \BibitemOpen
  \bibfield  {author} {\bibinfo {author} {\bibfnamefont {R.~G.}\ \bibnamefont {Chambers}},\ }\bibfield  {title} {\bibinfo {title} {The kinetic formulation of conduction problems},\ }\href {https://doi.org/10.1088/0370-1298/65/6/114} {\bibfield  {journal} {\bibinfo  {journal} {Proceedings of the Physical Society. Section A}\ }\textbf {\bibinfo {volume} {65}},\ \bibinfo {pages} {458} (\bibinfo {year} {1952})}\BibitemShut {NoStop}%
\bibitem [{\citenamefont {Fanfarillo}\ \emph {et~al.}(2009)\citenamefont {Fanfarillo}, \citenamefont {Benfatto}, \citenamefont {Caprara}, \citenamefont {Castellani},\ and\ \citenamefont {Grilli}}]{Fanfarillo_PRB2009}%
  \BibitemOpen
  \bibfield  {author} {\bibinfo {author} {\bibfnamefont {L.}~\bibnamefont {Fanfarillo}}, \bibinfo {author} {\bibfnamefont {L.}~\bibnamefont {Benfatto}}, \bibinfo {author} {\bibfnamefont {S.}~\bibnamefont {Caprara}}, \bibinfo {author} {\bibfnamefont {C.}~\bibnamefont {Castellani}},\ and\ \bibinfo {author} {\bibfnamefont {M.}~\bibnamefont {Grilli}},\ }\bibfield  {title} {\bibinfo {title} {Theory of fluctuation conductivity from interband pairing in pnictide superconductors},\ }\href {https://doi.org/10.1103/PhysRevB.79.172508} {\bibfield  {journal} {\bibinfo  {journal} {Phys. Rev. B}\ }\textbf {\bibinfo {volume} {79}},\ \bibinfo {pages} {172508} (\bibinfo {year} {2009})}\BibitemShut {NoStop}%
\bibitem [{\citenamefont {Fanfarillo}\ and\ \citenamefont {Benfatto}(2014)}]{Fanfarillo_SST2014}%
  \BibitemOpen
  \bibfield  {author} {\bibinfo {author} {\bibfnamefont {L.}~\bibnamefont {Fanfarillo}}\ and\ \bibinfo {author} {\bibfnamefont {L.}~\bibnamefont {Benfatto}},\ }\bibfield  {title} {\bibinfo {title} {Anisotropy of the superconducting fluctuations in multiband superconductors: the case of {L}i{F}e{A}s},\ }\href {https://doi.org/10.1088/0953-2048/27/12/124009} {\bibfield  {journal} {\bibinfo  {journal} {Superconductor Science and Technology}\ }\textbf {\bibinfo {volume} {27}},\ \bibinfo {pages} {124009} (\bibinfo {year} {2014})}\BibitemShut {NoStop}%
\bibitem [{\citenamefont {Dorin}\ \emph {et~al.}(1993)\citenamefont {Dorin}, \citenamefont {Klemm}, \citenamefont {Varlamov}, \citenamefont {Buzdin},\ and\ \citenamefont {Livanov}}]{Dorin_PRB1993}%
  \BibitemOpen
  \bibfield  {author} {\bibinfo {author} {\bibfnamefont {V.~V.}\ \bibnamefont {Dorin}}, \bibinfo {author} {\bibfnamefont {R.~A.}\ \bibnamefont {Klemm}}, \bibinfo {author} {\bibfnamefont {A.~A.}\ \bibnamefont {Varlamov}}, \bibinfo {author} {\bibfnamefont {A.~I.}\ \bibnamefont {Buzdin}},\ and\ \bibinfo {author} {\bibfnamefont {D.~V.}\ \bibnamefont {Livanov}},\ }\bibfield  {title} {\bibinfo {title} {Fluctuation conductivity of layered superconductors in a perpendicular magnetic field},\ }\href {https://doi.org/10.1103/PhysRevB.48.12951} {\bibfield  {journal} {\bibinfo  {journal} {Phys. Rev. B}\ }\textbf {\bibinfo {volume} {48}},\ \bibinfo {pages} {12951} (\bibinfo {year} {1993})}\BibitemShut {NoStop}%
\bibitem [{\citenamefont {Ovchenkov}\ \emph {et~al.}(2018)\citenamefont {Ovchenkov}, \citenamefont {Chareev}, \citenamefont {Kulbachinskii}, \citenamefont {Kytin}, \citenamefont {Presnov}, \citenamefont {Skourski}, \citenamefont {Volkova},\ and\ \citenamefont {Vasiliev}}]{Ovchenkov_JMMM2018}%
  \BibitemOpen
  \bibfield  {author} {\bibinfo {author} {\bibfnamefont {Y.}~\bibnamefont {Ovchenkov}}, \bibinfo {author} {\bibfnamefont {D.}~\bibnamefont {Chareev}}, \bibinfo {author} {\bibfnamefont {V.}~\bibnamefont {Kulbachinskii}}, \bibinfo {author} {\bibfnamefont {V.}~\bibnamefont {Kytin}}, \bibinfo {author} {\bibfnamefont {D.}~\bibnamefont {Presnov}}, \bibinfo {author} {\bibfnamefont {Y.}~\bibnamefont {Skourski}}, \bibinfo {author} {\bibfnamefont {O.}~\bibnamefont {Volkova}},\ and\ \bibinfo {author} {\bibfnamefont {A.}~\bibnamefont {Vasiliev}},\ }\bibfield  {title} {\bibinfo {title} {Magnetotransport properties of {F}e{S}e in fields up to 50 {T}},\ }\href {https://doi.org/https://doi.org/10.1016/j.jmmm.2017.10.108} {\bibfield  {journal} {\bibinfo  {journal} {Journal of Magnetism and Magnetic Materials}\ }\textbf {\bibinfo {volume} {459}},\ \bibinfo {pages} {221} (\bibinfo {year} {2018})},\ \bibinfo {note} {the selected papers of Seventh Moscow International Symposium on Magnetism (MISM-2017)}\BibitemShut {NoStop}%
\end{thebibliography}
\end{document}